\documentclass[preprint,superscriptaddress,nofootinbib]{revtex4}
  \usepackage{graphicx}
  \begin{document}
  \title{Study of the ${\Upsilon}(1S)$ ${\to}$ $B_{c}D_{s}^{\ast}$
         decay with pQCD approach}
  \author{Junfeng Sun}
  \affiliation{Institute of Particle and Nuclear Physics,
              Henan Normal University, Xinxiang 453007, China}
  \author{Yueling Yang}
  \affiliation{Institute of Particle and Nuclear Physics,
              Henan Normal University, Xinxiang 453007, China}
  \author{Qingxia Li}
  \affiliation{Institute of Particle and Nuclear Physics,
              Henan Normal University, Xinxiang 453007, China}
  \author{Gongru Lu}
  \affiliation{Institute of Particle and Nuclear Physics,
              Henan Normal University, Xinxiang 453007, China}
  \author{Jinshu Huang}
  \affiliation{College of Physics and Electronic Engineering,
              Nanyang Normal University, Nanyang 473061, China}
  \author{Qin Chang}
  \affiliation{Institute of Particle and Nuclear Physics,
              Henan Normal University, Xinxiang 453007, China}
  \begin{abstract}
  The ${\Upsilon}(1S)$ ${\to}$ $B_{c}D_{s}^{\ast}$ weak decay
  is studied with the perturbative QCD approach firstly.
  It is found that
  (1) main contributions to branching ratio come from the
  longitudinal and parallel helicity amplitudes,
  (2) branching ratio, longitudinal and parallel polarization
  fractions are sensitive to the wave functions of the
  ${\Upsilon}(1S)$ meson,
  (3) branching ratio for the ${\Upsilon}(1S)$ ${\to}$
  $B_{c}D_{s}^{\ast}$ decay can reach up to $10^{-9}$,
  which might be promisingly measured by the future
  experiments.
  \end{abstract}
  \maketitle

  \section{Introduction}
  \label{sec01}
  The ${\Upsilon}(1S)$ meson is the ground spin-triplet $S$-wave
  state of bottomonium (bound state consists of both the bottom
  quark $b$ and the anti-bottom quark $\bar{b}$) with the
  quantum number of $I^{G}J^{PC}$ $=$ $0^{-}1^{--}$ \cite{pdg}.
  The ${\Upsilon}(1S)$ meson lies below the kinematic open-bottom
  threshold. It is generally believed that the ${\Upsilon}(1S)$
  meson decays mainly through the annihilation of the $b\bar{b}$
  pairs into three gluons, one photon, two gluons plus one photon,
  with branching ratios \cite{pdg} of some $(81.7{\pm}0.7)\%$,
  $(3+R){\cal B}r_{{\ell}{\ell}}$, $(2.2{\pm}0.6)\%$,
  respectively, where $R$ ${\simeq}$ $10/3$
  is the ratio of the hadron production rate to the lepton
  ${\mu}^{+}{\mu}^{-}$ pair production rate at the
  energy scale of $m_{{\Upsilon}(1S)}$, and
  ${\cal B}r_{{\ell}{\ell}}$ is the branching ratio for
  the pure leptonic ${\Upsilon}(1S)$ ${\to}$
  ${\ell}^{+}{\ell}^{-}$ decay.
  One of the prominent features is its narrow width,
  ${\Gamma}_{{\Upsilon}(1S)}$ $=$ $54.02{\pm}1.25$ keV
  \cite{pdg}, while the ${\Upsilon}(1S)$ meson is about
  ten times heavier than the nucleon,
  $m_{{\Upsilon}(1S)}$ $=$ $9460.30{\pm}0.26$ MeV
  \cite{pdg}.
  This fact can be explained by the following argument.
  The hadronic decay ${\Upsilon}(1S)$ ${\to}$ $ggg$ is suppressed by the
  phenomenological Okubo-Zweig-Iizuka (OZI) rule \cite{o,z,i}.
  The branching ratio ${\cal B}r_{{\ell}{\ell}}$ is
  proportional (a) to the square of the electric charge
  of the bottom quark, $e_{b}^{2}$ where $e_{b}$ $=$ $-1/3$
  in the unit of ${\vert}e{\vert}$;
  (b) to the square of the electromagnetic coupling constant,
  ${\alpha}^{2}$ where ${\alpha}$ ${\approx}$ $1/128$;
  and (c) to the energy dependence of the photon propagator,
  $1/m_{{\Upsilon}(1S)}^{2}$ \cite{annu1993}.

  Besides the above-mentioned strong, electromagnetic and
  radiative decay mechanisms, the ${\Upsilon}(1S)$ meson
  can also decay via the weak interaction within the
  standard model. As it is well known, over $10^{8}$
  ${\Upsilon}(1S)$ data samples have been accumulated
  at Belle \cite{epjc74}. More and more upsilon data
  samples are hopefully expected at the running LHC and
  the forthcoming SuperKEKB.
  Although branching ratio for the ${\Upsilon}(1S)$ weak
  decay is tiny, about $2/{\tau}_{B}{\Gamma}_{{\Upsilon}(1S)}$
  ${\sim}$ ${\cal O}(10^{-8})$ where ${\tau}_{B}$ is the
  lifetime of the $B_{u,d}$ meson, there seems
  to exist a realistic possibility to search for the
  ${\Upsilon}(1S)$ weak decay at future experiments.
  In this paper, we will study the ${\Upsilon}(1S)$ ${\to}$
  $B_{c}D_{s}^{\ast}$ weak decay with the perturbative QCD
  (pQCD) approach \cite{pqcd1,pqcd2,pqcd3}.

  Experimentally, branching ratios for the two-body leptonic
  ${\Upsilon}(1S)$ decays, some OZI-suppressed hadronic
  ${\Upsilon}(1S)$ decays and radiative decays have been measured,
  but there is still no measurement reprot on the magnetic dipole transition
  decay ${\Upsilon}(1S)$ ${\to}$ ${\gamma}{\eta}_{b}$ and weak decays
  for the moment \cite{pdg}.
  The ${\Upsilon}(1S)$ weak decay posses a unique structure due to the
  Cabibbo-Kobayashi-Maskawa (CKM) matrix properties which predicts
  the channels with one $B_{c}^{(\ast)}$ meson are dominant.
  The signals for the ${\Upsilon}(1S)$
  ${\to}$ $B_{c}D_{s}^{\ast}$ weak decay should, in principle,
  be easily distinguished from possibly intricate background,
  due to the facts that the back-to-back final states with
  opposite electric charges have definite momentum and energy
  in the rest frame of the ${\Upsilon}(1S)$ meson.
  The identification of a single flavored either $B_{c}$ or
  $D_{s}^{\ast}$ meson could be used as an effective selection criterion.
  Moreover, the radiative decay of the $D_{s}^{\ast}$ meson
  can provide a useful extra signal and a powerful constraint.
  Of course, any evidences of an abnormally large branching
  ratio for the ${\Upsilon}(1S)$ weak decay might be a hint
  of new physics.

  Theoretically, in recent years, many attractive methods have been
  fully developed, such as the pQCD approach \cite{pqcd1,pqcd2,pqcd3},
  the QCD factorization (QCDF) approach \cite{qcdf1,qcdf2,qcdf3},
  soft and collinear effective theory \cite{scet1,scet2,scet3,scet4},
  and widely applied to accommodate measurements on the $B$ meson
  weak decays.
  The ${\Upsilon}(1S)$ weak decays permit one to cross
  check parameters obtained from the $B$ meson decay, and to test
  various phenomenological models.
  The ${\Upsilon}(1S)$ weak decays into final states containing one $B_{c}$
  meson are favorable processes due to the CKM factor $V_{cb}$,
  which also provide an additional occasion to scrutinize the underlying
  structure of doubly-heavy hadrons, and to improve our understanding
  on the short- and long-distance contributions in heavy quark weak decay.
  The semileptonic decays ${\Upsilon}(1S)$ ${\to}$
  $B_{c}{\ell}^{-}\bar{\nu}_{\ell}$ (${\ell}$ $=$ $e$, ${\mu}$, ${\tau}$)
  have been studied based on the Bauer-Stech-Wirbel model \cite{rohit}.
  The two-body nonleptonic decays ${\Upsilon}(1S)$ ${\to}$ $B_{c}M$
  ($M$ $=$ ${\pi}$, $K^{(\ast)}$, ${\rho}$) have been investigated recently by
  employing the factorization scheme, such as the naive factorization
  approximation \cite{rohit,ijmpa1999}, the QCD-improved QCDF formulation
  \cite{adv-sun,jpg-sun} and the pQCD approach \cite{prd92,plb751}.
  The ${\Upsilon}(1S)$ ${\to}$ $B_{c}D_{s}^{\ast}$ weak decay is
  favored by color and the CKM factor ${\vert}V_{cb}V_{cs}^{\ast}{\vert}$,
  so it should, in principle, have relatively large branching ratio
  among the ${\Upsilon}(1S)$ weak decays.
  However, there is still no theoretical study
  devoted to the ${\Upsilon}(1S)$ ${\to}$ $B_{c}D_{s}^{\ast}$
  weak decay now.
  In this paper, we will investigate the ${\Upsilon}(1S)$
  ${\to}$ $B_{c}D_{s}^{\ast}$ decay with the pQCD approach
  to offer a ready reference for the future experiments.

  This paper is organized as follows.
  The theoretical framework and the amplitudes for the
  ${\Upsilon}(1S)$ ${\to}$ $B_{c}D_{s}^{\ast}$ decay are
  presented in section \ref{sec02}.
  The numerical results and discussion are given in
  section \ref{sec03}. The last section is a summary.

  \section{theoretical framework}
  \label{sec02}
  \subsection{The effective Hamiltonian}
  \label{sec0201}
  The effective Hamiltonian responsible for the
  ${\Upsilon}(1S)$ ${\to}$ $B_{c}D_{s}^{\ast}$ weak
  decay is \cite{9512380}
   \begin{equation}
  {\cal H}_{\rm eff}\, =\,
   \frac{G_{F}}{\sqrt{2}}\,
   \Big\{ V_{cb} V_{cs}^{\ast}
   \sum\limits_{i=1}^{2} C_{i}({\mu})\,Q_{i}({\mu})
  -V_{tb} V_{ts}^{\ast}
   \sum\limits_{j=3}^{10}
   C_{j}({\mu})\,Q_{j}({\mu}) \Big\}
   + {\rm H.c.}
   \label{hamilton},
   \end{equation}
  where the Fermi coupling constant $G_{F}$ ${\simeq}$
  $1.166{\times}10^{-5}\,{\rm GeV}^{-2}$ \cite{pdg};
  with the Wolfenstein parameterization, the CKM factors
  are written as \cite{pdg},
  \begin{eqnarray}
  V_{cb}V_{cs}^{\ast} &=&
  +            A{\lambda}^{2}
  - \frac{1}{2}A{\lambda}^{4}
  - \frac{1}{8}A{\lambda}^{6}(1+4A^{2})
  +{\cal O}({\lambda}^{7})
  \label{eq:ckm01}, \\
  V_{tb}V_{ts}^{\ast} &=& -V_{cb}V_{cs}^{\ast}
  - A{\lambda}^{4}({\rho}-i{\eta})
  +{\cal O}({\lambda}^{7})
  \label{eq:ckm02}.
  \end{eqnarray}

  The local tree operators $Q_{1,2}$ and penguin operators
  $Q_{3,{\cdots},10}$ are defined below.
    \begin{eqnarray}
    Q_{1} &=&
  [ \bar{c}_{\alpha}{\gamma}_{\mu}(1-{\gamma}_{5})b_{\alpha} ]
  [ \bar{s}_{\beta} {\gamma}^{\mu}(1-{\gamma}_{5})c_{\beta} ]
    \label{q1}, \\
    Q_{2} &=&
  [ \bar{c}_{\alpha}{\gamma}_{\mu}(1-{\gamma}_{5})b_{\beta} ]
  [ \bar{s}_{\beta}{\gamma}^{\mu}(1-{\gamma}_{5})c_{\alpha} ]
    \label{q2},
    \end{eqnarray}
    \begin{eqnarray}
    Q_{3} &=& \sum\limits_{q}
  [ \bar{s}_{\alpha}{\gamma}_{\mu}(1-{\gamma}_{5})b_{\alpha} ]
  [ \bar{q}_{\beta} {\gamma}^{\mu}(1-{\gamma}_{5})q_{\beta} ]
    \label{q3}, \\
    Q_{4} &=& \sum\limits_{q}
  [ \bar{s}_{\alpha}{\gamma}_{\mu}(1-{\gamma}_{5})b_{\beta} ]
  [ \bar{q}_{\beta}{\gamma}^{\mu}(1-{\gamma}_{5})q_{\alpha} ]
    \label{q4}, \\
    Q_{5} &=& \sum\limits_{q}
  [ \bar{s}_{\alpha}{\gamma}_{\mu}(1-{\gamma}_{5})b_{\alpha} ]
  [ \bar{q}_{\beta} {\gamma}^{\mu}(1+{\gamma}_{5})q_{\beta} ]
    \label{q5}, \\
    Q_{6} &=& \sum\limits_{q}
  [ \bar{s}_{\alpha}{\gamma}_{\mu}(1-{\gamma}_{5})b_{\beta} ]
  [ \bar{q}_{\beta}{\gamma}^{\mu}(1+{\gamma}_{5})q_{\alpha} ]
    \label{q6},
    \end{eqnarray}
    \begin{eqnarray}
    Q_{7} &=& \sum\limits_{q} \frac{3}{2}e_{q}\,
  [ \bar{s}_{\alpha}{\gamma}_{\mu}(1-{\gamma}_{5})b_{\alpha} ]
  [ \bar{q}_{\beta} {\gamma}^{\mu}(1+{\gamma}_{5})q_{\beta} ]
    \label{q7}, \\
    Q_{8} &=& \sum\limits_{q} \frac{3}{2}e_{q}\,
  [ \bar{s}_{\alpha}{\gamma}_{\mu}(1-{\gamma}_{5})b_{\beta} ]
  [ \bar{q}_{\beta}{\gamma}^{\mu}(1+{\gamma}_{5})q_{\alpha} ]
    \label{q8}, \\
    Q_{9} &=& \sum\limits_{q} \frac{3}{2}e_{q}\,
  [ \bar{s}_{\alpha}{\gamma}_{\mu}(1-{\gamma}_{5})b_{\alpha} ]
  [ \bar{q}_{\beta} {\gamma}^{\mu}(1-{\gamma}_{5})q_{\beta} ]
    \label{q9}, \\
    Q_{10} &=& \sum\limits_{q} \frac{3}{2}e_{q}\,
  [ \bar{s}_{\alpha}{\gamma}_{\mu}(1-{\gamma}_{5})b_{\beta} ]
  [ \bar{q}_{\beta}{\gamma}^{\mu}(1-{\gamma}_{5})q_{\alpha} ]
    \label{q10},
    \end{eqnarray}
  where ${\alpha}$ and ${\beta}$ are color indices;
  $q$ denotes all the active quarks at the scale of
  ${\mu}$ ${\sim}$ ${\cal O}(m_{b})$, i.e.,
  $q$ $=$ $u$, $d$, $s$, $c$, $b$;
  and $e_{q}$ is the electric charge of the $q$ quark
  in the unit of ${\vert}e{\vert}$.

  The Wilson coefficients $C_{i}(\mu)$ summarize the physical
  contributions above the scale of ${\mu}$, and could be
  reliably calculated with the renormalization group improved
  perturbation theory. The physical contributions below the
  scale of ${\mu}$ are included in the hadronic matrix elements
  (HME) where the local operators sandwiched between initial
  and final hadron states.
  To obtain the decay amplitudes, the remaining work
  is to calculate HME properly.

  \subsection{Hadronic matrix elements}
  \label{sec0202}
  Phenomenologically, combining the $k_{T}$ factorization
  theorem \cite{npb366} with the collinear factorization
  hypothesis, and using the Lepage-Brodsky approach for
  exclusive processes \cite{prd22}, HME can be written
  as the convolution of universal wave functions reflecting
  the nonperturbative contributions with hard scattering
  subamplitudes containing the perturbative contributions
  within the pQCD framework, where the transverse momentum
  of valence quarks is retained and the Sudakov factor
  is introduced, in order to regulate the endpoint
  singularities and provide a naturally dynamical cutoff
  on the nonperturbative contributions \cite{pqcd1,pqcd2,pqcd3}.
  Generally, the decay amplitude can be separated into three
  parts: the Wilson coefficients $C_{i}$ incorporating the
  hard contributions above the typical scale of $t$,
  the process-dependent scattering amplitudes $T$ accounting
  for the heavy quark decay, and the universal wave functions
  ${\Phi}$ including the soft and long-distance contributions,
  i.e.,
  \begin{equation}
  {\int} dk\,
  C_{i}(t)\,T(t,k)\,{\Phi}(k)e^{-S}
  \label{hadronic},
  \end{equation}
  where $k$ is the momentum of valence quarks, and
  $e^{-S}$ is the Sudakov factor.

  \subsection{Kinematic variables}
  \label{sec0203}
  The light cone kinematic variables in the ${\Upsilon}(1S)$
  rest frame are defined as follows.
  \begin{equation}
  p_{{\Upsilon}(1S)}\, =\, p_{1}\, =\, \frac{m_{1}}{\sqrt{2}}(1,1,0)
  \label{kine-p1},
  \end{equation}
  \begin{equation}
  p_{B_{c}}\, =\, p_{2}\, =\, (p_{2}^{+},p_{2}^{-},0)
  \label{kine-p2},
  \end{equation}
  \begin{equation}
  p_{D_{s}^{\ast}}\, =\, p_{3}\, =\, (p_{3}^{-},p_{3}^{+},0)
  \label{kine-p3},
  \end{equation}
  \begin{equation}
  k_{i}\, =\, x_{i}\,p_{i}+(0,0,\vec{k}_{iT})
  \label{kine-ki},
  \end{equation}
  \begin{equation}
 {\epsilon}_{i}^{\parallel}\, =\,
  \frac{p_{i}}{m_{i}}-\frac{m_{i}}{p_{i}{\cdot}n_{+}}n_{+}
  \label{kine-longe},
  \end{equation}
  \begin{equation}
 {\epsilon}_{i}^{\perp}\, =\, (0,0,\vec{1})
  \label{kine-transe},
  \end{equation}
  \begin{equation}
  n_{+}=(1,0,0)
  \label{kine-null},
  \end{equation}
  \begin{equation}
  p_{i}^{\pm}\, =\, (E_{i}\,{\pm}\,p)/\sqrt{2}
  \label{kine-pipm},
  \end{equation}
  \begin{equation}
  s\, =\, 2\,p_{2}{\cdot}p_{3}
  \label{kine-s},
  \end{equation}
  \begin{equation}
  t\, =\, 2\,p_{1}{\cdot}p_{2}\, =\ 2\,m_{1}\,E_{2}
  \label{kine-t},
  \end{equation}
  \begin{equation}
  u\, =\, 2\,p_{1}{\cdot}p_{3}\, =\ 2\,m_{1}\,E_{3}
  \label{kine-u},
  \end{equation}
  \begin{equation}
  p = \frac{\sqrt{ [m_{1}^{2}-(m_{2}+m_{3})^{2}]\,[m_{1}^{2}-(m_{2}-m_{3})^{2}] }}{2\,m_{1}}
  \label{kine-pcm},
  \end{equation}
  where $x_{i}$ and $\vec{k}_{iT}$ are the longitudinal momentum
  fraction and transverse momentum of the valence quark, respectively;
  ${\epsilon}_{i}^{\parallel}$ and ${\epsilon}_{i}^{\perp}$ are the
  longitudinal and transverse polarization vectors, respectively,
  satisfying with the relations ${\epsilon}_{i}^{2}$ $=$ $-1$
  and ${\epsilon}_{i}{\cdot}p_{i}$ $=$ $0$;
  the subscript $i$ $=$ $1$, $2$, $3$ on variables $p_{i}$, $E_{i}$, $m_{i}$,
  and ${\epsilon}_{i}^{{\parallel},{\perp}}$ correspond to the
  ${\Upsilon}(1S)$, $B_{c}$, $D_{s}^{\ast}$ mesons, respectively;
  $n_{+}$ is the null vector; $s$, $t$ and $u$ are the Lorentz-invariant
  variables; $p$ is the common momentum of final states.
  The notation of momentum is displayed in Fig.\ref{fig:fey}(a).

  \subsection{Wave functions}
  \label{sec0204}
  With the notation in \cite{jhep0605,jhep0703}, the
  definitions of the diquark operator HME are
  \begin{equation}
 {\langle}0{\vert}b_{i}(z)\bar{b}_{j}(0){\vert}
 {\Upsilon}(p_{1},{\epsilon}_{1}^{{\parallel}}){\rangle}\,
 =\, \frac{f_{{\Upsilon}(1S)}}{4}{\int}dk_{1}\,e^{-ik_{1}{\cdot}z}
  \Big\{ \!\!\not{\epsilon}_{1}^{{\parallel}} \Big[
   m_{1}\,{\phi}_{\Upsilon}^{v}(k_{1})
  -\!\!\not{p}_{1}\, {\phi}_{\Upsilon}^{t}(k_{1})
  \Big] \Big\}_{ji}
  \label{wave-bbl},
  \end{equation}
  \begin{equation}
 {\langle}0{\vert}b_{i}(z)\bar{b}_{j}(0){\vert}
 {\Upsilon}(p_{1},{\epsilon}_{1}^{{\perp}}){\rangle}\,
 =\, \frac{f_{{\Upsilon}(1S)}}{4}{\int}dk_{1}\,e^{-ik_{1}{\cdot}z}
  \Big\{ \!\!\not{\epsilon}_{1}^{{\perp}} \Big[
   m_{1}\,{\phi}_{\Upsilon}^{V}(k_{1})
  -\!\!\not{p}_{1}\, {\phi}_{\Upsilon}^{T}(k_{1})
  \Big] \Big\}_{ji}
  \label{wave-bbt},
  \end{equation}
  \begin{equation}
 {\langle}B_{c}(p_{2}){\vert}\bar{c}_{i}(z)b_{j}(0){\vert}0{\rangle}\,
 =\, \frac{if_{B_{c}}}{4}{\int}dk_{2}\,e^{ik_{2}{\cdot}z}\,
  \Big\{ {\gamma}_{5}\Big[ \!\!\not{p}_{2}+m_{2}\Big]
 {\phi}_{B_{c}}(k_{2}) \Big\}_{ji}
  \label{wave-bc},
  \end{equation}
  \begin{equation}
 {\langle}D_{s}^{{\ast}}(p_{3},{\epsilon}_{3}^{{\parallel}})
 {\vert}c_{i}(0)\bar{s}_{j}(z){\vert}0{\rangle}\ =\
  \frac{f_{D_{s}^{\ast}}}{4}{\int}_{0}^{1}dk_{3}\,e^{ik_{3}{\cdot}z}
  \Big\{ \!\not{\epsilon}_{3}^{{\parallel}} \Big[
   m_{3}\,{\phi}_{D_{s}^{\ast}}^{v}(k_{3})
  +\!\not{p}_{3}\, {\phi}_{D_{s}^{\ast}}^{t}(k_{3})
  \Big] \Big\}_{ji}
  \label{wave-dslong},
  \end{equation}
  \begin{equation}
 {\langle}D_{s}^{{\ast}}(p_{3},{\epsilon}_{3}^{{\perp}})
 {\vert}c_{i}(0)\bar{s}_{j}(z){\vert}0{\rangle}\ =\
  \frac{f_{D_{s}^{\ast}}}{4}{\int}_{0}^{1}dk_{3}\,e^{ik_{3}{\cdot}z}
  \Big\{ \!\not{\epsilon}_{3}^{{\perp}} \Big[
   m_{3}\,{\phi}_{D_{s}^{\ast}}^{V}(k_{3})
 +\!\not{p}_{3}\, {\phi}_{D_{s}^{\ast}}^{T}(k_{3})
  \Big] \Big\}_{ji}
  \label{wave-dsperp},
  \end{equation}
  where $f_{{\Upsilon}(1S)}$, $f_{B_{c}}$, $f_{D_{s}^{\ast}}$ are
  decay constants.

  Because of the mass relations,
  $m_{{\Upsilon}(1S)}$ ${\simeq}$ $2m_{b}$,
  $m_{B_{c}}$ ${\simeq}$ $m_{b}$ $+$ $m_{c}$,
  and $m_{D_{s}^{\ast}}$ ${\simeq}$ $m_{c}$ $+$ $m_{s}$
  (see Table \ref{tab:input}),
  it might assume that the motion of the valence quarks
  in all participating mesons is nearly nonrelativistic.
  The wave functions of the ${\Upsilon}(1S)$, $B_{c}$, $D_{s}^{\ast}$
  mesons could be approximately described with the
  nonrelativistic quantum chromodynamics \cite{prd46,prd51,rmp77}
  and Schr\"{o}dinger equation.
  Combining the wave functions of a nonrelativistic isotropic
  harmonic oscillator potential with their asymptotic
  forms \cite{jhep0605,jhep0703}, we obtain \cite{prd92},
   \begin{equation}
  {\phi}_{\Upsilon}^{v}(x) = {\phi}_{\Upsilon}^{T}(x) = A\, x\bar{x}\,
  {\exp}\Big\{ -\frac{m_{b}^{2}}{8\,{\beta}_{1}^{2}\,x\,\bar{x}} \Big\}
   \label{wave-bblv},
   \end{equation}
   \begin{equation}
  {\phi}_{\Upsilon}^{t}(x) = B\, t^{2}\,
  {\exp}\Big\{ -\frac{m_{b}^{2}}{8\,{\beta}_{1}^{2}\,x\,\bar{x}} \Big\}
   \label{wave-bblt},
   \end{equation}
   \begin{equation}
  {\phi}_{\Upsilon}^{V}(x) = C\, (1+t^{2})\,
  {\exp}\Big\{ -\frac{m_{b}^{2}}{8\,{\beta}_{1}^{2}\,x\,\bar{x}} \Big\}
   \label{wave-bbtt},
   \end{equation}
   \begin{equation}
  {\phi}_{B_{c}}(x) = D\, x\bar{x}\,
  {\exp}\Big\{ -\frac{\bar{x}\,m_{c}^{2}+x\,m_{b}^{2}}
                     {8\,{\beta}_{2}^{2}\,x\,\bar{x}} \Big\}
   \label{wave-bcv},
   \end{equation}
   \begin{equation}
  {\phi}_{D_{s}^{\ast}}^{v}(x) = {\phi}_{D_{s}^{\ast}}^{T}(x) = E\, x\bar{x}\,
  {\exp}\Big\{ -\frac{\bar{x}\,m_{s}^{2}+x\,m_{c}^{2}}
                     {8\,{\beta}_{3}^{2}\,x\,\bar{x}} \Big\}
   \label{wave-dslv},
   \end{equation}
   \begin{equation}
  {\phi}_{D_{s}^{\ast}}^{t}(x) = F\, t^{2}\,
  {\exp}\Big\{ -\frac{\bar{x}\,m_{s}^{2}+x\,m_{c}^{2}}
                     {8\,{\beta}_{3}^{2}\,x\,\bar{x}} \Big\}
   \label{wave-dslt},
   \end{equation}
   \begin{equation}
  {\phi}_{D_{s}^{\ast}}^{V}(x) = G\, (1+t^{2})\,
  {\exp}\Big\{ -\frac{\bar{x}\,m_{s}^{2}+x\,m_{c}^{2}}
                     {8\,{\beta}_{3}^{2}\,x\,\bar{x}} \Big\}
   \label{wave-dstt},
   \end{equation}
   where $\bar{x}$ $=$ $1$ $-$ $x$; $t$ $=$ $x$ $-$ $\bar{x}$;
   ${\beta}_{i}$ $=$ ${\xi}_{i}\,{\alpha}_{s}({\xi}_{i})$
   with ${\xi}_{i}$ $=$ $m_{i}/2$ based on the NRQCD power
   counting rules \cite{prd46};
   ${\alpha}_{s}$ is the QCD coupling constant;
   the exponential function represents the $k_{T}$ distribution;
   parameters $A$, $B$, $C$, $D$, $E$, $F$, $G$ are the
   normalization coefficients satisfying the conditions
   \begin{equation}
  {\int}_{0}^{1}dx\,{\phi}_{B_{c}}(x)=1
   \label{wave-cbc},
   \end{equation}
   \begin{equation}
  {\int}_{0}^{1}dx\,{\phi}_{\Upsilon}^{i}(x) = 1,
   \quad \text{for}\ \ i=v,t,V,T
   \label{wave-cbb},
   \end{equation}
   \begin{equation}
  {\int}_{0}^{1}dx\,{\phi}_{D_{s}^{\ast}}^{i}(x) = 1,
   \quad \text{for}\ \ i=v,t,V,T
   \label{wave-cds}.
   \end{equation}
  \begin{figure}[h]
  \includegraphics[width=0.98\textwidth,bb=80 610 530 710]{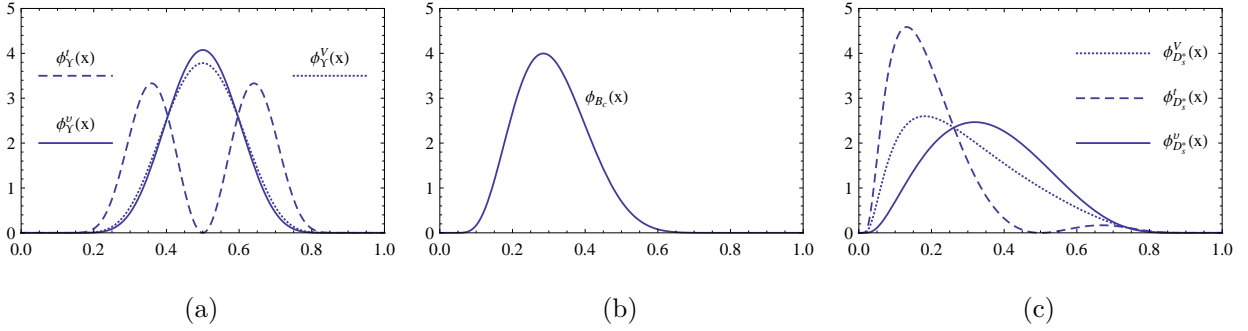}
  \caption{The distribution amplitudes for the
   ${\Upsilon}(1S)$, $B_{c}$, $D_{s}^{\ast}$ mesons in (a), (b), (c),
   respectively.}
  \label{fig:wave}
  \end{figure}

  The shape lines of the normalized distribution amplitudes for
  the ${\Upsilon}(1S)$, $B_{c}$, $D_{s}^{\ast}$ mesons are
  showed in Fig.\ref{fig:wave}.
  It is clearly seen that
  (1) distribution amplitudes for the ${\Upsilon}(1S)$,
  $B_{c}$, $D_{s}^{\ast}$ mesons shrink rapidly to zero
  at the endpoint $x$ ${\to}$ $0$, $1$ due to the suppression
  from the exponential functions,
  (2) although the nonrelativistic model of wave functions
  is crude, distribution amplitudes Eq.(\ref{wave-bblv})-Eq.(\ref{wave-dstt}) can
  reflect, at least to some extent, the feature that the
  valence quarks share momentum fractions
  according to their masses.

  \subsection{Decay amplitudes}
  \label{sec0205}
\begin{figure}[h]
  \includegraphics[width=0.95\textwidth,bb=80 530 530 720]{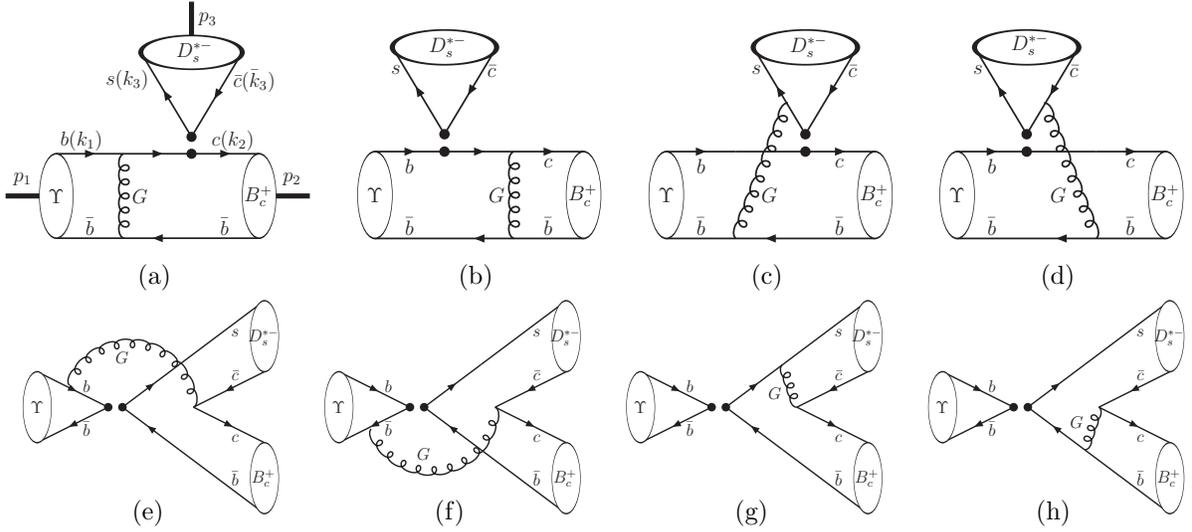}
  \caption{Feynman diagrams for the ${\Upsilon}(1S)$ ${\to}$
  $B_{c}D_{s}^{\ast}$ decay with the pQCD approach, where
  (a,b) are the factorizable emission diagrams,
  (c,d) are the nonfactorizable emission diagrams,
  (e,f) are the nonfactorizable annihilation diagrams,
  and (g,h) are the factorizable annihilation diagrams.}
  \label{fig:fey}
  \end{figure}

  The Feynman diagrams for the ${\Upsilon}(1S)$ ${\to}$
  $B_{c}D_{s}^{\ast}$ weak decay are shown in Fig.\ref{fig:fey}.
  There are two types. One is the emission topology, and the
  other is annihilation topology. Each type is further
  subdivided into factorizable diagram where gluon attaches
  to quarks in the same meson, and nonfactorizable diagrams
  where gluon connects to quarks between different mesons.

  The amplitude for the ${\Upsilon}(1S)$ ${\to}$ $B_{c}D_{s}^{\ast}$
  weak decay is defined as below \cite{prd66},
   \begin{equation}
  {\cal A}({\Upsilon}(1S){\to}B_{c}D_{s}^{\ast})\ =\
  {\cal A}_{L}({\epsilon}_{1}^{{\parallel}},{\epsilon}_{3}^{{\parallel}})
 +{\cal A}_{N}({\epsilon}_{1}^{{\perp}},{\epsilon}_{3}^{{\perp}})
 +i\,{\cal A}_{T}\,{\varepsilon}_{{\mu}{\nu}{\alpha}{\beta}}\,
  {\epsilon}_{1}^{{\mu}}\,{\epsilon}_{3}^{{\nu}}\,
   p_{1}^{\alpha}\,p_{3}^{\beta}
   \label{eq:amp01},
   \end{equation}
  which is conventionally written as the helicity amplitudes \cite{prd66},
   \begin{equation}
  {\cal A}_{0}\ =\ -C_{\cal A}\,
  {\cal A}_{L}({\epsilon}_{1}^{{\parallel}},{\epsilon}_{3}^{{\parallel}})
   \label{eq:amp02},
   \end{equation}
   \begin{equation}
  {\cal A}_{\parallel}\ =\ \sqrt{2}\,C_{\cal A}
  {\cal A}_{N}({\epsilon}_{1}^{{\perp}},{\epsilon}_{3}^{{\perp}})
   \label{eq:amp03},
   \end{equation}
   \begin{equation}
  {\cal A}_{\perp}\ =\ \sqrt{2}\,C_{\cal A}\,m_{1}\,p\, {\cal A}_{T}
   \label{eq:amp04},
   \end{equation}
   \begin{equation}
  C_{\cal A}\ =\ i\frac{G_{F}}{\sqrt{2}}\,\frac{C_{F}}{N_{c}}\,
  {\pi}\, f_{{\Upsilon}(1S)}\,f_{B_{c}}\, f_{D_{s}^{\ast}}
   \label{eq:amp05},
   \end{equation}
  and the polarization amplitude ${\cal A}_{j}$ is written as
   \begin{eqnarray}
  {\cal A}_{j} &=&
   V_{cb} V_{cs}^{\ast}\,\Big\{
   \Big( {\cal A}_{a,j}^{LL}+{\cal A}_{b,j}^{LL} \Big) a_{1}
  +\Big( {\cal A}_{c,j}^{LL}+{\cal A}_{d,j}^{LL} \Big) C_{2}
   \Big\}
   \nonumber \\ &-&
   V_{tb} V_{tq}^{\ast}\, \Big\{
   \Big( {\cal A}_{a,j}^{LL}+{\cal A}_{b,j}^{LL} \Big)\,(a_{4}+a_{10})
  +\Big( {\cal A}_{c,j}^{LL}+{\cal A}_{d,j}^{LL} \Big)\, (C_{3}+C_{9})
   \nonumber \\ & & \quad
  +\Big( {\cal A}_{e,j}^{LL}+{\cal A}_{f,j}^{LL} \Big)\,
  (C_{3}+C_{4}-\frac{1}{2}C_{9}-\frac{1}{2}C_{10})
   \nonumber \\ & & \quad
  +\Big( {\cal A}_{g,j}^{LL}+{\cal A}_{h,j}^{LL} \Big)\,
  (a_{3}+a_{4}-\frac{1}{2}a_{9}-\frac{1}{2}a_{10})
   \nonumber \\ & & \quad
  +\Big( {\cal A}_{e,j}^{LR}+{\cal A}_{f,j}^{LR} \Big)\,
   (C_{6}-\frac{1}{2}C_{8})
  +\Big( {\cal A}_{g,j}^{LR}+{\cal A}_{h,j}^{LR} \Big)\,
   (a_{5}-\frac{1}{2}a_{7})
   \nonumber \\ & & \quad
  +\Big( {\cal A}_{c,j}^{SP}+{\cal A}_{d,j}^{SP} \Big)\,
   (C_{5}+C_{7})
  +\Big( {\cal A}_{e,j}^{SP}+{\cal A}_{f,j}^{SP} \Big)\,
   (C_{5}-\frac{1}{2}C_{7}) \Big\}
   \label{amp-all},
   \end{eqnarray}
  where $C_{F}$ $=$ $4/3$ and the color number $N_{c}$ $=$ $3$;
  for the building blocks ${\cal A}_{i,j}^{k}$,
  the first subscript $i$ corresponds to the indices of Fig.\ref{fig:fey};
  the second subscript $j$ $=$ $L$, $N$, $T$ denotes to three
  different helicity amplitudes;
  the superscript $k$ refers to three possible Dirac structures
  ${\Gamma}_{1}{\otimes}{\Gamma}_{2}$ of the four-quark operator
  $(\bar{q}_{1}{\Gamma}_{1}q_{2})(\bar{q}_{1}{\Gamma}_{2}q_{2})$,
  namely $k$ $=$ $LL$ for $(V-A){\otimes}(V-A)$,
  $k$ $=$ $LR$ for $(V-A){\otimes}(V+A)$, and
  $k$ $=$ $SP$ for $-2(S-P){\otimes}(S+P)$;
  The explicit expressions of building blocks ${\cal A}_{i,j}^{k}$
  are collected in Appendix \ref{blocks}.
  The parameter $a_{i}$ is defined as follows.
  \begin{equation}
  a_{i} = \left\{ \begin{array}{l}
  C_{i}+C_{i+1}/N_{c}, \quad \text{for odd $i$}; \\
  C_{i}+C_{i-1}/N_{c}, \quad \text{for even $i$}.
  \end{array} \right.
  \label{eq:ai}
  \end{equation}

  \section{Numerical results and discussion}
  \label{sec03}
  In the rest frame of the ${\Upsilon}(1S)$ meson,
  branching ratio (${\cal B}r$),
  polarization fractions ($f_{0,{\parallel},{\perp}}$)
  and relative phases (${\phi}_{{\parallel},{\perp}}$)
  between helicity amplitudes (${\cal A}_{0,{\parallel},{\perp}}$)
  for the ${\Upsilon}(1S)$
  ${\to}$ $B_{c}D_{s}^{\ast}$ weak decay are defined as
   \begin{equation}
  {\cal B}r\ =\ \frac{1}{12{\pi}}\,
   \frac{p}{m_{{\Upsilon}(1S)}^{2}{\Gamma}_{{\Upsilon}(1S)}}\, \Big\{
  {\vert}{\cal A}_{0}{\vert}^{2}+{\vert}{\cal A}_{\parallel}{\vert}^{2}
 +{\vert}{\cal A}_{\perp}{\vert}^{2} \Big\}
   \label{br},
   \end{equation}
   \begin{equation}
  f_{0,{\parallel},{\perp}}\ =\
   \frac{ {\vert}{\cal A}_{0,{\parallel},{\perp}}{\vert}^{2} }{
  {\vert}{\cal A}_{0}{\vert}^{2}+{\vert}{\cal A}_{\parallel}{\vert}^{2}
 +{\vert}{\cal A}_{\perp}{\vert}^{2} }
   \label{f0},
   \end{equation}
   \begin{equation}
  {\phi}_{{\parallel},{\perp}}\ =\ {\arg} (
   {\cal A}_{{\parallel},{\perp}} / {\cal A}_{0} )
   \label{phi}.
   \end{equation}

   \begin{table}[h]
   \caption{The numerical values of input parameters.}
   \label{tab:input}
   \begin{ruledtabular}
   \begin{tabular}{ll}
   \multicolumn{2}{c}{The Wolfenstein parameters\footnotemark[1]} \\ \hline
     $A$ $=$ $0.814^{+0.023}_{-0.024}$ \cite{pdg},
   & ${\lambda}$ $=$ $0.22537{\pm}0.00061$ \cite{pdg}, \\
     $\bar{\rho}$ $=$ $0.117{\pm}0.021$ \cite{pdg},
   & $\bar{\eta}$ $=$ $0.353{\pm}0.013$ \cite{pdg}, \\ \hline
   \multicolumn{2}{c}{Mass, width and decay constant} \\ \hline
     $m_{b}$ $=$ $4.78{\pm}0.06$ GeV \cite{pdg},
   & $m_{c}$ $=$ $1.67{\pm}0.07$ GeV \cite{pdg}, \\
     $m_{s}$ ${\simeq}$ $510$ MeV \cite{book},
   & ${\Gamma}_{{\Upsilon}(1S)}$ $=$ $54.02{\pm}1.25$ keV \cite{pdg}, \\
     $m_{{\Upsilon}(1S)}$ $=$ $9460.30{\pm}0.26$ MeV \cite{pdg}
   & $f_{{\Upsilon}(1S)}$ $=$ $676.4{\pm}10.7$ MeV \cite{prd92} \\
     $m_{B_{c}}$ $=$ $6275.6{\pm}1.1$ MeV \cite{pdg},
   & $f_{B_{c}}$ $=$ $427{\pm}6$ MeV \cite{fbc}, \\
     $m_{D_{s}^{\ast}}$ $=$ $2112.1{\pm}0.4$ MeV \cite{pdg},
   & $f_{D_{s}^{\ast}}$ $=$ $274{\pm}6$ MeV \cite{fds}.
   \end{tabular}
   \end{ruledtabular}
   \footnotetext[1]{The relation between parameters (${\rho}$, ${\eta}$)
   and ($\bar{\rho}$, $\bar{\eta}$) is \cite{pdg}: $({\rho}+i{\eta})$ $=$
    $\displaystyle \frac{ \sqrt{1-A^{2}{\lambda}^{4}}(\bar{\rho}+i\bar{\eta}) }
    { \sqrt{1-{\lambda}^{2}}[1-A^{2}{\lambda}^{4}(\bar{\rho}+i\bar{\eta})] }$.}
   \end{table}

  The input parameters are listed in Table \ref{tab:input}.
  If not specified explicitly, we will take their central
  values as the default inputs.
  Our numerical results are collected in Table. \ref{tab:case},
  where the first uncertainty comes from the CKM parameters,
  the second uncertainty is from the choice of the typical
  scale $(1{\pm}0.1)t_{i}$ and $t_{i}$ is given in
  Eqs.(\ref{tab}-\ref{tgh});
  the third uncertainty is from the variation of
  mass $m_{b}$ and $m_{c}$.
  The following are some comments.

   \begin{table}[h]
   \caption{Branching ratio, polarization fractions, and relative
   phases for different cases, where we use the wave functions of
   Eqs.(\ref{wave-bblv}-\ref{wave-bbtt}) and
   Eqs.(\ref{wave-dslv}-\ref{wave-dstt}) for the ${\Upsilon}(1S)$
   and $D_{s}^{\ast}$ meson, respectively in case A; we use the
   same wave functions for both the transversal and longitudinal
   polarization ${\Upsilon}(1S)$ meson in case B, i.e.,
   ${\phi}_{\Upsilon}^{v,t,V,T}$ = Eq.(\ref{wave-bblv});
   case C for ${\phi}_{D_{s}^{\ast}}^{v,t,V,T}$ = Eq.(\ref{wave-dslv});
   case D for ${\phi}_{\Upsilon}^{v,t,V,T}$ = Eq.(\ref{wave-bblv})
   and ${\phi}_{D_{s}^{\ast}}^{v,t,V,T}$ = Eq.(\ref{wave-dslv}).}
   \label{tab:case}
   \begin{ruledtabular}
   \begin{tabular}{ccccc}
    & case A & case B & case C & case D \\ \hline
      $10^{9}{\times}{\cal B}r$
    & $1.68^{+0.12+0.24+0.10}_{-0.11-0.11-0.22}$
    & $2.10^{+0.14+0.29+0.09}_{-0.14-0.14-0.25}$
    & $1.63^{+0.11+0.18+0.10}_{-0.11-0.08-0.20}$
    & $2.04^{+0.14+0.22+0.08}_{-0.13-0.10-0.24}$ \\
      $10^{2}{\times}f_{0}$
    & $35.6^{+ 0.0+ 0.2+ 1.3}_{- 0.0- 0.1- 1.4}$
    & $47.8^{+ 0.0+ 0.1+ 0.2}_{- 0.0- 0.1- 0.1}$
    & $35.8^{+ 0.0+ 0.1+ 1.3}_{- 0.0- 0.0- 1.4}$
    & $48.0^{+ 0.0+ 0.1+ 0.2}_{- 0.0- 0.0- 0.1}$ \\
      $10^{2}{\times}f_{\parallel}$
    & $56.5^{+ 0.0+ 0.1+ 0.9}_{- 0.0- 0.2- 0.9}$
    & $46.0^{+ 0.0+ 0.0+ 0.1}_{- 0.0- 0.1- 0.5}$
    & $56.3^{+ 0.0+ 0.0+ 0.9}_{- 0.0- 0.1- 0.9}$
    & $45.8^{+ 0.0+ 0.0+ 0.1}_{- 0.0- 0.1- 0.5}$ \\
      $10^{2}{\times}f_{\perp}$
    & $ 7.9^{+ 0.0+ 0.0+ 0.5}_{- 0.0- 0.1- 0.4}$
    & $ 6.2^{+ 0.0+ 0.0+ 0.2}_{- 0.0- 0.0- 0.2}$
    & $ 7.8^{+ 0.0+ 0.0+ 0.5}_{- 0.0- 0.0- 0.4}$
    & $ 6.2^{+ 0.0+ 0.0+ 0.2}_{- 0.0- 0.0- 0.2}$ \\
      ${\phi}_{\parallel}$
    & ${\simeq}$ $1.9^{\circ}$
    & ${\simeq}$ $0.4^{\circ}$
    & ${\simeq}$ $2.1^{\circ}$
    & ${\simeq}$ $0.4^{\circ}$ \\
      ${\phi}_{\perp}$
    & ${\simeq}$ $-174.0^{\circ}$
    & ${\simeq}$ $-175.6^{\circ}$
    & ${\simeq}$ $-173.7^{\circ}$
    & ${\simeq}$ $-175.2^{\circ}$
   \end{tabular}
   \end{ruledtabular}
   \end{table}

  (1)
  Branching ratio for the ${\Upsilon}(1S)$ ${\to}$ $B_{c}D_{s}^{\ast}$
  decay can reach up to ${\cal O}(10^{-9})$ with the pQCD approach,
  which might be promisingly measurable at the running LHC and
  forthcoming SuperKEKB.
  For example, the ${\Upsilon}(1S)$ production cross section in
  p-Pb collision is about a few ${\mu}b$ at the LHCb \cite{jhep1407}
  and ALICE \cite{plb740} detectors. So, more than $10^{12}$ ${\Upsilon}(1S)$
  data samples could be in principle available per $ab^{-1}$ data
  collected by the LHCb and ALICE detectors, corresponding to a few thousands
  of the ${\Upsilon}(1S)$ ${\to}$ $B_{c}D_{s}^{\ast}$ events.

  (2) The contributions to branching ratio mainly come from the
  longitudinal and parallel polarization helicity amplitudes,
  $f_{0}$ $+$ $f_{{\parallel}}$ ${\gtrsim}$ 90\%, while the
  perpendicular polarization fraction $f_{{\perp}}$ is
  generally less than 10\%.
  From Table.\ref{tab:case}, it is seen that the polarization
  fractions are not sensitive to the input parameters.
  However, we find that branching ratio, polarization fractions
  $f_{0}$ and $f_{{\parallel}}$, are sensitive to the
  ${\Upsilon}(1S)$ wave functions (see Table.\ref{tab:case}).
  This might imply that the polarization measurement on
  the ${\Upsilon}(1S)$ ${\to}$ $B_{c}D_{s}^{\ast}$ decay
  would provide some information on the wave functions
  and thus the interquark binding forces responsible for
  the ${\Upsilon}(1S)$ meson.

  (3)
  The relative phase ${\phi}_{{\parallel}}$ is very small.
  This is consistent with prediction of the QCD factorization
  approach \cite{qcdf1,qcdf2}, where the strong phase arising
  from nonfactorizable contributions is suppressed by color
  and ${\alpha}_{s}$ for the $a_{1}$-dominated processes.
  The relative phases, if they could be determined experimentally,
  will improve our understanding on nonfactorizable contributions,
  factorization mechanism, and the strong dynamics at different
  energy scales.

  \begin{figure}[h]
  \includegraphics[width=0.5\textwidth]{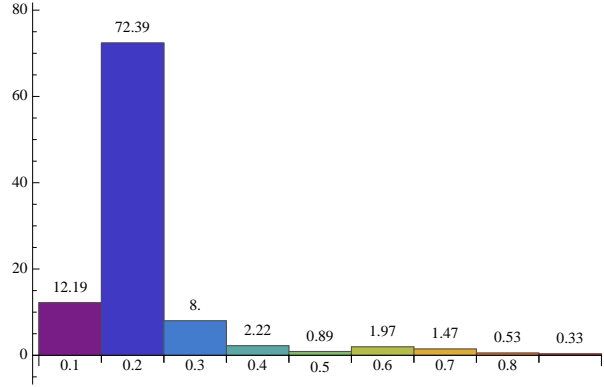}
  \caption{The contributions to the branching ratio from different
  region of ${\alpha}_{s}/{\pi}$ (horizontal axises), where the
  numbers over histogram denote the percentage of the corresponding
  contributions.}
  \label{fig:br-as}
  \end{figure}

  (4)
  As it is well known, due to the large mass of final states,
  the momentum transition in the ${\Upsilon}(1S)$ ${\to}$
  $B_{c}D_{s}^{\ast}$ decay may be not large enough. One might
  naturally wonder whether the pQCD approach is applicable
  and whether the perturbative calculation is reliable.
  Therefore, it is necessary to check what percentage
  of the contributions comes from the perturbative region.
  The contributions to branching ratio from different region
  of ${\alpha}_{s}/{\pi}$ are showed in Fig.\ref{fig:br-as}.
  It can be clearly seen that more than 90\% contributions
  to branching ratio come from the ${\alpha}_{s}/{\pi}$
  ${\le}$ $0.3$ region, implying that the calculation with
  the pQCD approach is reliable.
  As the discussion in \cite{pqcd1,pqcd2,pqcd3},
  there are many factors for this, for example,
  the choice of the typical scale in Eqs.(\ref{tab}-\ref{tgh}),
  retaining the quark transverse moment and introducing the
  Sudakov factor to suppress the nonperturbative contributions,
  which deserve much attention and further investigation, but
  beyond the scope of this paper.

  (5)
  Besides the uncertainties listed in Table \ref{tab:case},
  the decay constants, $f_{{\Upsilon}(1S)}$,  $f_{B_{c}}$,
  and $f_{D_{s}^{\ast}}$, can bring about 6\% uncertainties
  to branching ratios.
  Other factors, such as the models of wave functions,
  contributions of higher order corrections to HME,
  relativistic effects, and so on, deserve
  the dedicated study. Our results just provide an order
  of magnitude estimation.

  \section{Summary}
  \label{sec04}
  The ${\Upsilon}(1S)$ weak decay is allowable within the
  standard model.
  With anticipation of the potential prospects of the
  ${\Upsilon}(1S)$ physics at high-luminosity dedicated
  heavy-flavor factories,
  the ${\Upsilon}(1S)$ ${\to}$ $B_{c}D_{s}^{\ast}$
  weak decay is studied with the pQCD approach firstly.
  It is found that
  (1) the longitudinal plus parallel
  polarization fractions are main shares, but sensitive
  to the ${\Upsilon}(1S)$ wave functions;
  (2) branching ratio for the ${\Upsilon}(1S)$ ${\to}$
  $B_{c}D_{s}^{\ast}$ weak decay can reach up to ${\cal O}(10^{-9})$,
  which might be measurable at the future experiments.

  \section*{Acknowledgments}
  We thank Professor Dongsheng Du (IHEP@CAS) and Professor
  Yadong Yang (CCNU) for helpful discussion.
  The work is supported by the National Natural Science Foundation
  of China (Grant Nos. 11547014, 11475055, U1332103 and 11275057).

  \begin{appendix}
  \section{The building blocks of decay amplitudes}
  \label{blocks}
  For the sake of simplicity,
  we decompose the decay amplitude Eq.(\ref{amp-all})
  into some building blocks ${\cal A}_{i,j}^{k}$, where
  the subscript $i$ corresponds to the indices of Fig.\ref{fig:fey};
  the subscript $j$ relates with different helicity amplitudes;
  the superscript $k$ refers to one of the three possible Dirac
  structures ${\Gamma}_{1}{\otimes}{\Gamma}_{2}$ of the
  four-quark operator
  $(\bar{q}_{1}{\Gamma}_{1}q_{2})(\bar{q}_{1}{\Gamma}_{2}q_{2})$,
  namely
  $k$ $=$ $LL$ for $(V-A){\otimes}(V-A)$,
  $k$ $=$ $LR$ for $(V-A){\otimes}(V+A)$, and
  $k$ $=$ $SP$ for $-2(S-P){\otimes}(S+P)$.
  The explicit expressions of ${\cal A}_{i,j}^{k}$
  are written as follows.
   \begin{eqnarray}
  {\cal A}_{a,L}^{LL} &=&
  {\int}_{0}^{1}dx_{1} {\int}_{0}^{1}dx_{2}
  {\int}_{0}^{\infty}b_{1} db_{1}
  {\int}_{0}^{\infty}b_{2} db_{2}\,
  {\phi}_{\Upsilon}^{v}(x_{1})\,
   \nonumber \\ & &
  {\phi}_{B_{c}}(x_{2})\, E_{a}(t_{a})\,
   H_{ab}({\alpha}_{e},{\beta}_{a},b_{1},b_{2})\,
  {\alpha}_{s}(t_{a})
   \nonumber \\ & &
   \Big\{ m_{1}^{2}\,s+m_{2}\,m_{b}\,u-
   (4\,m_{1}^{2}\,p^{2}+m_{2}^{2}\,u)\,\bar{x}_{2} \Big\}
   \label{figaL-LL},
   \end{eqnarray}
   \begin{eqnarray}
  {\cal A}_{a,N}^{LL} &=& m_{1}\, m_{3}
  {\int}_{0}^{1}dx_{1} {\int}_{0}^{1}dx_{2}
  {\int}_{0}^{\infty}b_{1} db_{1}
  {\int}_{0}^{\infty}b_{2} db_{2}
   \nonumber \\ & &
  {\phi}_{\Upsilon}^{V}(x_{1})\,
  {\phi}_{B_{c}}(x_{2})\, E_{a}(t_{a})\,
   H_{ab}({\alpha}_{e},{\beta}_{a},b_{1},b_{2})
   \nonumber \\ & &
  {\alpha}_{s}(t_{a})\,
   \Big\{  2\,m_{2}^{2}\,\bar{x}_{2} -2\,m_{2}\,m_{b} -t \Big\}
   \label{figaN-LL},
   \end{eqnarray}
  \begin{eqnarray}
  {\cal A}_{a,T}^{LL} &=& 2\,m_{1}\, m_{3}
  {\int}_{0}^{1}dx_{1} {\int}_{0}^{1}dx_{2}
  {\int}_{0}^{\infty}b_{1} db_{1}
  {\int}_{0}^{\infty}b_{2} db_{2}
   \nonumber \\ & & \!\!\!\!\!\!\!\!\!\!\!\!
  {\phi}_{\Upsilon}^{V}(x_{1})\,
  {\phi}_{B_{c}}(x_{2})\, E_{a}(t_{a})\,
   H_{ab}({\alpha}_{e},{\beta}_{a},b_{1},b_{2})\,
  {\alpha}_{s}(t_{a})
   \label{figaT-LL},
   \end{eqnarray}
   \begin{eqnarray}
  {\cal A}_{b,L}^{LL} &=&
  {\int}_{0}^{1}dx_{1} {\int}_{0}^{1}dx_{2}
  {\int}_{0}^{\infty}b_{1} db_{1}
  {\int}_{0}^{\infty}b_{2} db_{2}
   \nonumber \\ & &
  {\phi}_{B_{c}}(x_{2})\, E_{b}(t_{b})\,
   H_{ab}({\alpha}_{e},{\beta}_{b},b_{2},b_{1})\,
  {\alpha}_{s}(t_{b})
   \nonumber \\ & &
   \Big\{ {\phi}_{\Upsilon}^{v}(x_{1})\, \Big[m_{1}^{2}\,
   (s-4\,p^{2})\,\bar{x}_{1}+2\,m_{2}\,m_{c}\,u-m_{2}^{2}\,u \Big]
   \nonumber \\ & &
   + {\phi}_{\Upsilon}^{t}(x_{1})\, m_{1}\,\Big[
   s\,(2\,m_{2}-m_{c})-2\,m_{2}\,u\,\bar{x}_{1} \Big] \Big\}
   \label{figbL-LL},
   \end{eqnarray}
   \begin{eqnarray}
  {\cal A}_{b,N}^{LL} &=& m_{3}
  {\int}_{0}^{1}dx_{1} {\int}_{0}^{1}dx_{2}
  {\int}_{0}^{\infty}b_{1} db_{1}
  {\int}_{0}^{\infty}b_{2} db_{2}
   \nonumber \\ & &
  {\phi}_{B_{c}}(x_{2})\, E_{b}(t_{b})\,
   H_{ab}({\alpha}_{e},{\beta}_{b},b_{2},b_{1})\,
  {\alpha}_{s}(t_{b})
   \nonumber \\ & &
   \Big\{ {\phi}_{\Upsilon}^{V}(x_{1})\,m_{1}\,
   \Big[ 2\,m_{2}^{2} -4\,m_{2}\,m_{c} -t\,\bar{x}_{1} \Big]
   \nonumber \\ & &
   + {\phi}_{\Upsilon}^{T}(x_{1})\, \Big[
   t\,(m_{c}-2\,m_{2})+4\,m_{1}^{2}\,m_{2}\,\bar{x}_{1} \Big]
   \Big\}
   \label{figbN-LL},
   \end{eqnarray}
   \begin{eqnarray}
  {\cal A}_{b,T}^{LL} &=& -2\,m_{3}
  {\int}_{0}^{1}dx_{1} {\int}_{0}^{1}dx_{2}
  {\int}_{0}^{\infty}b_{1} db_{1}
  {\int}_{0}^{\infty}b_{2} db_{2}
   \nonumber \\ & &
  {\phi}_{B_{c}}(x_{2})\, E_{b}(t_{b})\,
   H_{ab}({\alpha}_{e},{\beta}_{b},b_{2},b_{1})\,
  {\alpha}_{s}(t_{b})
   \nonumber \\ & &
   \Big\{ {\phi}_{\Upsilon}^{V}(x_{1})\,m_{1}\,\bar{x}_{1}
  + {\phi}_{\Upsilon}^{T}(x_{1})\, (m_{c}-2\,m_{2}) \Big\}
   \label{figbT-LL},
   \end{eqnarray}
   \begin{eqnarray}
  {\cal A}_{c,L}^{LL} &=& \frac{1}{N_{c}}
  {\int}_{0}^{1}dx_{1} {\int}_{0}^{1}dx_{2} {\int}_{0}^{1}dx_{3}
  {\int}_{0}^{\infty}db_{1}  {\int}_{0}^{\infty}b_{2}db_{2}
  {\int}_{0}^{\infty}b_{3}db_{3}
   \nonumber \\ & &
  {\phi}_{B_{c}}(x_{2})\, {\phi}_{D_{s}^{\ast}}^{v}(x_{3})\,
   E_{c}(t_{c})\, H_{cd}({\alpha}_{e},{\beta}_{c},b_{2},b_{3})\,
   {\alpha}_{s}(t_{c})\,
   \nonumber \\ & &
  {\delta}(b_{1}-b_{2})\,
   \Big\{ {\phi}_{\Upsilon}^{v}(x_{1})\,u\,
   \Big[ t\,x_{1}-2\,m_{2}^{2}\,x_{2}-s\,\bar{x}_{3} \Big]
   \nonumber \\ & &
   + {\phi}_{\Upsilon}^{t}(x_{1})\, m_{1}\,m_{2}\,
   \Big[ s\,x_{2}+2\,m_{3}^{2}\,\bar{x}_{3}-u\,x_{1}\Big] \Big\}
   \label{figcL-LL},
   \end{eqnarray}
   \begin{eqnarray}
  {\cal A}_{c,N}^{LL} &=& \frac{m_{3}}{N_{c}}
  {\int}_{0}^{1}dx_{1} {\int}_{0}^{1}dx_{2} {\int}_{0}^{1}dx_{3}
  {\int}_{0}^{\infty}db_{1}  {\int}_{0}^{\infty}b_{2}db_{2}
  {\int}_{0}^{\infty}b_{3}db_{3}
   \nonumber \\ & &
  {\phi}_{B_{c}}(x_{2})\, {\phi}_{D_{s}^{\ast}}^{V}(x_{3})\,
   E_{c}(t_{c})\, H_{cd}({\alpha}_{e},{\beta}_{c},b_{2},b_{3})\,
   {\alpha}_{s}(t_{c})\,
   \nonumber \\ & &
  {\delta}(b_{1}-b_{2})
   \Big\{ {\phi}_{\Upsilon}^{V}(x_{1})\,2\,m_{1}\,
   \Big[ 2\,m_{2}^{2}\,x_{2}+s\,\bar{x}_{3}-t\,x_{1} \Big]
   \nonumber \\ & &
   + {\phi}_{\Upsilon}^{T}(x_{1})\,m_{2}\,
   \Big[ 2\,m_{1}^{2}\,x_{1} -t\,x_{2}-u\,\bar{x}_{3} \Big]
   \Big\}
   \label{figcN-LL},
   \end{eqnarray}
   \begin{eqnarray}
  {\cal A}_{c,T}^{LL} &=& \frac{2\,m_{2}\,m_{3}}{N_{c}}
  {\int}_{0}^{1}dx_{1} {\int}_{0}^{1}dx_{2} {\int}_{0}^{1}dx_{3}
  {\int}_{0}^{\infty}db_{1}  {\int}_{0}^{\infty}b_{2}db_{2}
   \nonumber \\ & &
  {\int}_{0}^{\infty}b_{3}db_{3}\, {\delta}(b_{1}-b_{2})\,
  {\phi}_{\Upsilon}^{T}(x_{1})\, {\phi}_{B_{c}}(x_{2})\,
  {\phi}_{D_{s}^{\ast}}^{V}(x_{3})
   \nonumber \\ & &
   E_{c}(t_{c})\, H_{cd}({\alpha}_{e},{\beta}_{c},b_{2},b_{3})\,
   {\alpha}_{s}(t_{c})\, (x_{2}-\bar{x}_{3})
   \label{figcT-LL},
   \end{eqnarray}
   \begin{eqnarray}
  {\cal A}_{c,L}^{SP} &=& \frac{m_{3}}{N_{c}}
  {\int}_{0}^{1}dx_{1} {\int}_{0}^{1}dx_{2} {\int}_{0}^{1}dx_{3}
  {\int}_{0}^{\infty}db_{1}  {\int}_{0}^{\infty}b_{2}db_{2}
  {\int}_{0}^{\infty}b_{3}db_{3}
   \nonumber \\ & &
  {\phi}_{B_{c}}(x_{2})\, {\phi}_{D_{s}^{\ast}}^{t}(x_{3})\,
   E_{c}(t_{c})\, H_{cd}({\alpha}_{e},{\beta}_{c},b_{2},b_{3})\,
   {\alpha}_{s}(t_{c})\,
   \nonumber \\ & &
  {\delta}(b_{1}-b_{2})\,
   \Big\{ {\phi}_{\Upsilon}^{v}(x_{1})\,m_{2}\,
   \Big[ 2\,m_{1}^{2}\,x_{1}-t\,x_{2}-u\,\bar{x}_{3} \Big]
   \nonumber \\ & &
   + {\phi}_{\Upsilon}^{t}(x_{1})\, m_{1}\,
   \Big[ 2\,m_{2}^{2}\,x_{2}+s\,\bar{x}_{3}-t\,x_{1} \Big] \Big\}
   \label{figcL-SP},
   \end{eqnarray}
   \begin{eqnarray}
  {\cal A}_{c,N}^{SP} &=& \frac{1}{N_{c}}
  {\int}_{0}^{1}dx_{1} {\int}_{0}^{1}dx_{2} {\int}_{0}^{1}dx_{3}
  {\int}_{0}^{\infty}db_{1}  {\int}_{0}^{\infty}b_{2}db_{2}
  {\int}_{0}^{\infty}b_{3}db_{3}
   \nonumber \\ & &
  {\phi}_{B_{c}}(x_{2})\, {\phi}_{D_{s}^{\ast}}^{T}(x_{3})\,
   E_{c}(t_{c})\, H_{cd}({\alpha}_{e},{\beta}_{c},b_{2},b_{3})\,
   {\alpha}_{s}(t_{c})\,
   \nonumber \\ & &
  {\delta}(b_{1}-b_{2})\,
   \Big\{ {\phi}_{\Upsilon}^{V}(x_{1})\,m_{1}\,m_{2}\,
   \Big[ s\,x_{2}+2\,m_{3}^{2}\,\bar{x}_{3}-u\,x_{1} \Big]
   \nonumber \\ & &
   + {\phi}_{\Upsilon}^{T}(x_{1})\, \Big[ m_{1}^{2}\,s\,x_{1}
   +(m_{2}^{2}\,u-s\,t)\,x_{2}-m_{3}^{2}\,t\,\bar{x}_{3} \Big] \Big\}
   \label{figcN-SP},
   \end{eqnarray}
   \begin{eqnarray}
  {\cal A}_{c,T}^{SP} &=& \frac{1}{N_{c}}
  {\int}_{0}^{1}dx_{1} {\int}_{0}^{1}dx_{2} {\int}_{0}^{1}dx_{3}
  {\int}_{0}^{\infty}db_{1}  {\int}_{0}^{\infty}b_{2}db_{2}
  {\int}_{0}^{\infty}b_{3}db_{3}
   \nonumber \\ & &
  {\phi}_{B_{c}}(x_{2})\, {\phi}_{D_{s}^{\ast}}^{T}(x_{3})\,
   E_{c}(t_{c})\, H_{cd}({\alpha}_{e},{\beta}_{c},b_{2},b_{3})\,
   {\alpha}_{s}(t_{c})\,
   \nonumber \\ & &
   \Big\{ {\phi}_{\Upsilon}^{T}(x_{1})\, \Big[
   (s+t)\,x_{2}+ 2\,m_{3}^{2}\,\bar{x}_{3}
   -(t+u)\,x_{1} \Big]
   \nonumber \\ & &
   + {\phi}_{\Upsilon}^{V}(x_{1})\,2\,m_{1}\,m_{2}\,(x_{1}-x_{2})
   \Big\}\,{\delta}(b_{1}-b_{2})
   \label{figcT-SP},
   \end{eqnarray}
   \begin{eqnarray}
  {\cal A}_{d,L}^{LL} &=& \frac{1}{N_{c}}
  {\int}_{0}^{1}dx_{1} {\int}_{0}^{1}dx_{2} {\int}_{0}^{1}dx_{3}
  {\int}_{0}^{\infty}db_{1}  {\int}_{0}^{\infty}b_{2}db_{2}
  {\int}_{0}^{\infty}b_{3}db_{3}
   \nonumber \\ & &
  {\phi}_{B_{c}}(x_{2})\, E_{d}(t_{d})\,
  H_{cd}({\alpha}_{e},{\beta}_{d},b_{2},b_{3})\,
   {\alpha}_{s}(t_{d})\, {\delta}(b_{1}-b_{2})
   \nonumber \\ & &
   \Big\{ {\phi}_{\Upsilon}^{t}(x_{1})\,
  {\phi}_{D_{s}^{\ast}}^{v}(x_{3})\,
  m_{1}\,m_{2}\,\Big[s\,x_{2}+2\,m_{3}^{2}\,x_{3}-u\,x_{1} \Big]
   \nonumber \\ & &
 +{\phi}_{\Upsilon}^{v}(x_{1})\,
  {\phi}_{D_{s}^{\ast}}^{v}(x_{3})\,
   4\,m_{1}^{2}\,p^{2}\,(x_{3}-x_{2})
   \nonumber \\ & &
 -{\phi}_{\Upsilon}^{v}(x_{1})\,
  {\phi}_{D_{s}^{\ast}}^{t}(x_{3})\,
   m_{3}\,m_{c}\,t \Big\}
   \label{figdL-LL},
   \end{eqnarray}
   \begin{eqnarray}
  {\cal A}_{d,N}^{LL} &=& \frac{1}{N_{c}}
  {\int}_{0}^{1}dx_{1} {\int}_{0}^{1}dx_{2} {\int}_{0}^{1}dx_{3}
  {\int}_{0}^{\infty}db_{1}  {\int}_{0}^{\infty}b_{2}db_{2}
  {\int}_{0}^{\infty}b_{3}db_{3}
   \nonumber \\ & &
  {\phi}_{B_{c}}(x_{2})\, E_{d}(t_{d})\,
  H_{cd}({\alpha}_{e},{\beta}_{d},b_{2},b_{3})\,
   {\alpha}_{s}(t_{d})\, {\delta}(b_{1}-b_{2})
   \nonumber \\ & &
   \Big\{ {\phi}_{\Upsilon}^{T}(x_{1})\,
  {\phi}_{D_{s}^{\ast}}^{V}(x_{3})\, m_{2}\,m_{3}\,
  \Big[ 2\,m_{1}^{2}\,x_{1}-t\,x_{2}\,-u\,x_{3} \Big]
   \nonumber \\ & &
 +{\phi}_{\Upsilon}^{V}(x_{1})\,
  {\phi}_{D_{s}^{\ast}}^{T}(x_{3})\,
  m_{1}\,m_{c}\,s \Big\}
   \label{figdN-LL},
   \end{eqnarray}
   \begin{eqnarray}
  {\cal A}_{d,T}^{LL} &=& \frac{1}{N_{c}}
  {\int}_{0}^{1}dx_{1} {\int}_{0}^{1}dx_{2} {\int}_{0}^{1}dx_{3}
  {\int}_{0}^{\infty}db_{1}  {\int}_{0}^{\infty}b_{2}db_{2}
  {\int}_{0}^{\infty}b_{3}db_{3}
   \nonumber \\ & &
  {\phi}_{B_{c}}(x_{2})\, E_{d}(t_{d})\,
  H_{cd}({\alpha}_{e},{\beta}_{d},b_{2},b_{3})\,
   {\alpha}_{s}(t_{d})\, {\delta}(b_{1}-b_{2})
   \nonumber \\ & &
   \Big\{ {\phi}_{\Upsilon}^{T}(x_{1})\,
  {\phi}_{D_{s}^{\ast}}^{V}(x_{3})\, 2\,m_{2}\,m_{3}\,
  (x_{2}-x_{3})
   \nonumber \\ & &
 -{\phi}_{\Upsilon}^{V}(x_{1})\,
  {\phi}_{D_{s}^{\ast}}^{T}(x_{3})\,
   2\,m_{1}\,m_{c} \Big\}
   \label{figdT-LL},
   \end{eqnarray}
   \begin{eqnarray}
  {\cal A}_{d,L}^{SP} &=& \frac{1}{N_{c}}
  {\int}_{0}^{1}dx_{1} {\int}_{0}^{1}dx_{2} {\int}_{0}^{1}dx_{3}
  {\int}_{0}^{\infty}db_{1}  {\int}_{0}^{\infty}b_{2}db_{2}
  {\int}_{0}^{\infty}b_{3}db_{3}
   \nonumber \\ & &
  {\phi}_{B_{c}}(x_{2})\, E_{d}(t_{d})\,
  H_{cd}({\alpha}_{e},{\beta}_{d},b_{2},b_{3})\,
   {\alpha}_{s}(t_{d})\, {\delta}(b_{1}-b_{2})
   \nonumber \\ & &
   \Big\{ {\phi}_{D_{s}^{\ast}}^{v}(x_{3})\, m_{c}
   \Big[ {\phi}_{\Upsilon}^{t}(x_{1})\,m_{1}\,s
   - {\phi}_{\Upsilon}^{v}(x_{1})\,m_{2}\,u \Big]
   \nonumber \\ & &
 +{\phi}_{\Upsilon}^{v}(x_{1})\,
  {\phi}_{D_{s}^{\ast}}^{t}(x_{3})\,
  m_{2}\,m_{3} \Big[2\,m_{1}^{2}\,x_{1}-t\,x_{2}-u\,x_{3}\Big]
   \nonumber \\ & &
 +{\phi}_{\Upsilon}^{t}(x_{1})\,
  {\phi}_{D_{s}^{\ast}}^{t}(x_{3})\,
  m_{1}\,m_{3} \Big[2\,m_{2}^{2}\,x_{2}+s\,x_{3}-t\,x_{1}\Big] \Big\}
   \label{figdL-SP},
   \end{eqnarray}
   \begin{eqnarray}
  {\cal A}_{d,N}^{SP} &=& \frac{1}{N_{c}}
  {\int}_{0}^{1}dx_{1} {\int}_{0}^{1}dx_{2} {\int}_{0}^{1}dx_{3}
  {\int}_{0}^{\infty}db_{1}  {\int}_{0}^{\infty}b_{2}db_{2}
  {\int}_{0}^{\infty}b_{3}db_{3}
   \nonumber \\ & &
  {\phi}_{B_{c}}(x_{2})\, E_{d}(t_{d})\,
  H_{cd}({\alpha}_{e},{\beta}_{d},b_{2},b_{3})\,
   {\alpha}_{s}(t_{d})\, {\delta}(b_{1}-b_{2})
   \nonumber \\ & &
   \Big\{ {\phi}_{D_{s}^{\ast}}^{V}(x_{3})\, m_{3}\,m_{c}
   \Big[ {\phi}_{\Upsilon}^{V}(x_{1})\,2\,m_{1}\,m_{2}
   - {\phi}_{\Upsilon}^{T}(x_{1})\,t \Big]
   \nonumber \\ & &
 +{\phi}_{D_{s}^{\ast}}^{T}(x_{3}) \Big[
  {\phi}_{\Upsilon}^{V}(x_{1})\,
  m_{1}\,m_{2}\,( s\,x_{2}+2\,m_{3}^{2}\,x_{3}-u\,x_{1} )
   \nonumber \\ & &
 +{\phi}_{\Upsilon}^{T}(x_{1})\,
   \{ m_{1}^{2}\,s\,x_{1}+(m_{2}^{2}\,u-s\,t)\,x_{2}-m_{3}^{2}\,t\,x_{3}
   \} \Big] \Big\}
   \label{figdN-SP},
   \end{eqnarray}
   \begin{eqnarray}
  {\cal A}_{d,T}^{SP} &=& \frac{1}{N_{c}}
  {\int}_{0}^{1}dx_{1} {\int}_{0}^{1}dx_{2} {\int}_{0}^{1}dx_{3}
  {\int}_{0}^{\infty}db_{1}  {\int}_{0}^{\infty}b_{2}db_{2}
  {\int}_{0}^{\infty}b_{3}db_{3}
   \nonumber \\ & &
  {\phi}_{B_{c}}(x_{2})\, E_{d}(t_{d})\,
  H_{cd}({\alpha}_{e},{\beta}_{d},b_{2},b_{3})\,
   {\alpha}_{s}(t_{d})\, {\delta}(b_{1}-b_{2})
   \nonumber \\ & &
   \Big\{ {\phi}_{D_{s}^{\ast}}^{T}(x_{3})
   \Big[ {\phi}_{\Upsilon}^{V}(x_{1})\,
   2\,m_{1}\,m_{2}\,(x_{1}-x_{2})
   \nonumber \\ & &
  +{\phi}_{\Upsilon}^{T}(x_{1})\, \{
  2\,m_{3}^{2}\,x_{3}+(s+t)\,x_{2}-(u+t)\,x_{1} \} \Big]
   \nonumber \\ & &
  + {\phi}_{\Upsilon}^{T}(x_{1})\,
   {\phi}_{D_{s}^{\ast}}^{V}(x_{3})\,
   2\,m_{3}\,m_{c} \Big\}
   \label{figdT-SP},
   \end{eqnarray}
   \begin{eqnarray}
  {\cal A}_{e,L}^{LL} &=& \frac{1}{N_{c}}
  {\int}_{0}^{1}dx_{1} {\int}_{0}^{1}dx_{2} {\int}_{0}^{1}dx_{3}
  {\int}_{0}^{\infty}b_{1}db_{1}  {\int}_{0}^{\infty}b_{2}db_{2}
  {\int}_{0}^{\infty}db_{3}
   \nonumber \\ & &
  {\phi}_{B_{c}}(x_{2})\, E_{e}(t_{e})\,
  H_{ef}({\alpha}_{a},{\beta}_{e},b_{1},b_{2})\,
   {\alpha}_{s}(t_{e})\, {\delta}(b_{2}-b_{3})
   \nonumber \\ & &
   \Big\{ {\phi}_{\Upsilon}^{v}(x_{1})\, \Big[
  {\phi}_{D_{s}^{\ast}}^{v}(x_{3})\,u\,
   \Big( t\,x_{1}-2\,m_{2}^{2}\,x_{2}-s\,\bar{x}_{3} \Big)
   \nonumber \\ & &
 +{\phi}_{D_{s}^{\ast}}^{t}(x_{3})\,m_{2}\,m_{3}\,
   \Big( 2\,m_{1}^{2}\,x_{1}-t\,x_{2}-u\,\bar{x}_{3}
   \Big) \Big]
   \nonumber \\ & &
 -{\phi}_{\Upsilon}^{t}(x_{1})\, m_{1}\,m_{b} \, \Big[
  {\phi}_{D_{s}^{\ast}}^{v}(x_{3})\,s+
  {\phi}_{D_{s}^{\ast}}^{t}(x_{3})\,4\,m_{2}\,m_{3}
   \Big] \Big\}
   \label{figeL-LL},
   \end{eqnarray}
   \begin{eqnarray}
  {\cal A}_{e,N}^{LL} &=& \frac{1}{N_{c}}
  {\int}_{0}^{1}dx_{1} {\int}_{0}^{1}dx_{2} {\int}_{0}^{1}dx_{3}
  {\int}_{0}^{\infty}b_{1}db_{1}  {\int}_{0}^{\infty}b_{2}db_{2}
  {\int}_{0}^{\infty}db_{3}
   \nonumber \\ & &
  {\phi}_{B_{c}}(x_{2})\, E_{e}(t_{e})\,
  H_{ef}({\alpha}_{a},{\beta}_{e},b_{1},b_{2})\,
   {\alpha}_{s}(t_{e})\, {\delta}(b_{2}-b_{3})
   \nonumber \\ & &
   \Big\{ {\phi}_{\Upsilon}^{V}(x_{1})\, \Big[
  {\phi}_{D_{s}^{\ast}}^{V}(x_{3})\,2\,m_{1}\,m_{3}\,
   \Big( 2\,m_{2}^{2}\,x_{2}+s\,\bar{x}_{3}-t\,x_{1} \Big)
   \nonumber \\ & &
 +{\phi}_{D_{s}^{\ast}}^{T}(x_{3})\,m_{1}\,m_{2}\,
   \Big( s\,x_{2}+2\,m_{3}^{2}\,\bar{x}_{3}-u\,x_{1}
   \Big) \Big]
   \nonumber \\ & &
 +{\phi}_{\Upsilon}^{T}(x_{1})\, m_{b} \, \Big[
  {\phi}_{D_{s}^{\ast}}^{V}(x_{3})\,m_{3}\,t+
  {\phi}_{D_{s}^{\ast}}^{T}(x_{3})\,2\,m_{2}\,u
   \Big] \Big\}
   \label{figeN-LL},
   \end{eqnarray}
   \begin{eqnarray}
  {\cal A}_{e,T}^{LL} &=& \frac{1}{N_{c}}
  {\int}_{0}^{1}dx_{1} {\int}_{0}^{1}dx_{2} {\int}_{0}^{1}dx_{3}
  {\int}_{0}^{\infty}b_{1}db_{1}  {\int}_{0}^{\infty}b_{2}db_{2}
  {\int}_{0}^{\infty}db_{3}
   \nonumber \\ & &
  {\phi}_{B_{c}}(x_{2})\, E_{e}(t_{e})\,
  H_{ef}({\alpha}_{a},{\beta}_{e},b_{1},b_{2})\,
   {\alpha}_{s}(t_{e})\, {\delta}(b_{2}-b_{3})
   \nonumber \\ & &
   \Big\{ {\phi}_{\Upsilon}^{V}(x_{1})\,
  {\phi}_{D_{s}^{\ast}}^{T}(x_{3})\,2\,m_{1}\,m_{2}\,
  (x_{1}-x_{2})
   \nonumber \\ & &
 +{\phi}_{\Upsilon}^{T}(x_{1})\, 2\,m_{b} \, \Big[
  {\phi}_{D_{s}^{\ast}}^{V}(x_{3})\,m_{3} -
  {\phi}_{D_{s}^{\ast}}^{T}(x_{3})\,2\,m_{2}
   \Big] \Big\}
   \label{figeT-LL},
   \end{eqnarray}
   \begin{eqnarray}
  {\cal A}_{e,L}^{LR} &=& \frac{1}{N_{c}}
  {\int}_{0}^{1}dx_{1} {\int}_{0}^{1}dx_{2} {\int}_{0}^{1}dx_{3}
  {\int}_{0}^{\infty}b_{1}db_{1}  {\int}_{0}^{\infty}b_{2}db_{2}
  {\int}_{0}^{\infty}db_{3}
   \nonumber \\ & &
  {\phi}_{B_{c}}(x_{2})\, E_{e}(t_{e})\,
  H_{ef}({\alpha}_{a},{\beta}_{e},b_{1},b_{2})\,
   {\alpha}_{s}(t_{e})\, {\delta}(b_{2}-b_{3})
   \nonumber \\ & &
   \Big\{ {\phi}_{D_{s}^{\ast}}^{v}(x_{3})\, \Big[
  {\phi}_{\Upsilon}^{v}(x_{1})\,4\,m_{1}^{2}\,p^{2}\,(x_{2}-x_{1})
 +{\phi}_{\Upsilon}^{t}(x_{1})\,m_{1}\,m_{b}\,s \Big]
   \nonumber \\ & &
 +{\phi}_{\Upsilon}^{v}(x_{1})\,
  {\phi}_{D_{s}^{\ast}}^{t}(x_{3})\,m_{2}\,m_{3} \,
   \Big[ t\,x_{2}+u\,\bar{x}_{3}-2\,m_{1}^{2}\,x_{1}
   \Big] \Big\}
   \label{figeL-LR},
   \end{eqnarray}
   \begin{eqnarray}
  {\cal A}_{e,N}^{LR} &=& \frac{1}{N_{c}}
  {\int}_{0}^{1}dx_{1} {\int}_{0}^{1}dx_{2} {\int}_{0}^{1}dx_{3}
  {\int}_{0}^{\infty}b_{1}db_{1}  {\int}_{0}^{\infty}b_{2}db_{2}
  {\int}_{0}^{\infty}db_{3}
   \nonumber \\ & &
  {\phi}_{B_{c}}(x_{2})\, E_{e}(t_{e})\,
  H_{ef}({\alpha}_{a},{\beta}_{e},b_{1},b_{2})\,
   {\alpha}_{s}(t_{e})\, {\delta}(b_{2}-b_{3})
   \nonumber \\ & &
   \Big\{ {\phi}_{\Upsilon}^{V}(x_{1})\,
  {\phi}_{D_{s}^{\ast}}^{T}(x_{3})\, m_{1}\,m_{2}\,
   \Big[ u\,x_{1}-s\,x_{2}-2\,m_{3}^{2}\,\bar{x}_{3} \Big]
   \nonumber \\ & &
 -{\phi}_{\Upsilon}^{T}(x_{1})\,
  {\phi}_{D_{s}^{\ast}}^{V}(x_{3})\,m_{3}\,m_{b}\,t \Big\}
   \label{figeN-LR},
   \end{eqnarray}
   \begin{eqnarray}
  {\cal A}_{e,T}^{LR} &=& \frac{1}{N_{c}}
  {\int}_{0}^{1}dx_{1} {\int}_{0}^{1}dx_{2} {\int}_{0}^{1}dx_{3}
  {\int}_{0}^{\infty}b_{1}db_{1}  {\int}_{0}^{\infty}b_{2}db_{2}
  {\int}_{0}^{\infty}db_{3}
   \nonumber \\ & &
  {\phi}_{B_{c}}(x_{2})\, E_{e}(t_{e})\,
  H_{ef}({\alpha}_{a},{\beta}_{e},b_{1},b_{2})\,
   {\alpha}_{s}(t_{e})\, {\delta}(b_{2}-b_{3})
   \nonumber \\ & &
   \Big\{ {\phi}_{\Upsilon}^{V}(x_{1})\,
  {\phi}_{D_{s}^{\ast}}^{T}(x_{3})\, 2\,m_{1}\,m_{2}\,
  (x_{2}-x_{1})
   \nonumber \\ & &
 +{\phi}_{\Upsilon}^{T}(x_{1})\,
  {\phi}_{D_{s}^{\ast}}^{V}(x_{3})\,2\,m_{3}\,m_{b} \Big\}
   \label{figeT-LR},
   \end{eqnarray}
   \begin{eqnarray}
  {\cal A}_{e,L}^{SP} &=& \frac{1}{N_{c}}
  {\int}_{0}^{1}dx_{1} {\int}_{0}^{1}dx_{2} {\int}_{0}^{1}dx_{3}
  {\int}_{0}^{\infty}b_{1}db_{1}  {\int}_{0}^{\infty}b_{2}db_{2}
  {\int}_{0}^{\infty}db_{3}
   \nonumber \\ & &
  {\phi}_{B_{c}}(x_{2})\, E_{e}(t_{e})\,
  H_{ef}({\alpha}_{a},{\beta}_{e},b_{1},b_{2})\,
   {\alpha}_{s}(t_{e})\, {\delta}(b_{2}-b_{3})
   \nonumber \\ & &
   \Big\{ {\phi}_{\Upsilon}^{t}(x_{1})\,
  {\phi}_{D_{s}^{\ast}}^{v}(x_{3})\,m_{1}\,m_{2}\,
  (u\,x_{1}-s\,x_{2}-2\,m_{3}^{2}\,\bar{x}_{3})
   \nonumber \\ & &
  +{\phi}_{\Upsilon}^{t}(x_{1})\,
   {\phi}_{D_{s}^{\ast}}^{t}(x_{3})\,m_{1}\,m_{3}\,
   (t\,x_{1}-2\,m_{2}^{2}\,x_{2}-s\,\bar{x}_{3})
   \nonumber \\ & &
   -{\phi}_{\Upsilon}^{v}(x_{1})\, m_{b}\, \Big[
  {\phi}_{D_{s}^{\ast}}^{v}(x_{3})\,m_{2}\,u
 +{\phi}_{D_{s}^{\ast}}^{t}(x_{3})\,m_{3}\,t \Big] \Big\}
   \label{figeL-SP},
   \end{eqnarray}
   \begin{eqnarray}
  {\cal A}_{e,N}^{SP} &=& \frac{1}{N_{c}}
  {\int}_{0}^{1}dx_{1} {\int}_{0}^{1}dx_{2} {\int}_{0}^{1}dx_{3}
  {\int}_{0}^{\infty}b_{1}db_{1}  {\int}_{0}^{\infty}b_{2}db_{2}
  {\int}_{0}^{\infty}db_{3}
   \nonumber \\ & &
  {\phi}_{B_{c}}(x_{2})\, E_{e}(t_{e})\,
  H_{ef}({\alpha}_{a},{\beta}_{e},b_{1},b_{2})\,
   {\alpha}_{s}(t_{e})\, {\delta}(b_{2}-b_{3})
   \nonumber \\ & &
   \Big\{ {\phi}_{\Upsilon}^{V}(x_{1})\, m_{1}\,m_{b}\,
   \Big[ {\phi}_{D_{s}^{\ast}}^{V}(x_{3})\,2\,m_{2}\,m_{3}
  + {\phi}_{D_{s}^{\ast}}^{T}(x_{3})\,s \Big]
   \nonumber \\ & &
  +{\phi}_{\Upsilon}^{T}(x_{1})\, \Big[
   {\phi}_{D_{s}^{\ast}}^{V}(x_{3})\,m_{2}\,m_{3}\,
   (t\,x_{2}+u\,\bar{x}_{3}-2\,m_{1}^{2}\,x_{1})
   \nonumber \\ & &
  +{\phi}_{D_{s}^{\ast}}^{T}(x_{3})\, \{
   (s\,t-m_{2}^{2}\,u)\,x_{2}+m_{3}^{2}\,t\,\bar{x}_{3}
   -m_{1}^{2}\,s\,x_{1} \} \Big] \Big\}
   \label{figeN-SP},
   \end{eqnarray}
   \begin{eqnarray}
  {\cal A}_{e,T}^{SP} &=& \frac{1}{N_{c}}
  {\int}_{0}^{1}dx_{1} {\int}_{0}^{1}dx_{2} {\int}_{0}^{1}dx_{3}
  {\int}_{0}^{\infty}b_{1}db_{1}  {\int}_{0}^{\infty}b_{2}db_{2}
  {\int}_{0}^{\infty}db_{3}
   \nonumber \\ & &
  {\phi}_{B_{c}}(x_{2})\, E_{e}(t_{e})\,
  H_{ef}({\alpha}_{a},{\beta}_{e},b_{1},b_{2})\,
   {\alpha}_{s}(t_{e})\, {\delta}(b_{2}-b_{3})
   \nonumber \\ & &
   \Big\{ {\phi}_{\Upsilon}^{T}(x_{1})\, \Big[
   {\phi}_{D_{s}^{\ast}}^{V}(x_{3})\,2\,m_{2}\,m_{3}\,
   (x_{2}-\bar{x}_{3})
   \nonumber \\ & &
 +{\phi}_{D_{s}^{\ast}}^{T}(x_{3}) \{
  (s+t)\,x_{2}+(u-s)\,\bar{x}_{3}-2\,m_{1}^{2}\,x_{1} \} \Big]
   \nonumber \\ & &
 +{\phi}_{\Upsilon}^{V}(x_{1})\,
  {\phi}_{D_{s}^{\ast}}^{T}(x_{3})\,2\,m_{1}\,m_{b} \Big\}
   \label{figeT-SP},
   \end{eqnarray}
   \begin{eqnarray}
  {\cal A}_{f,L}^{LL} &=& \frac{1}{N_{c}}
  {\int}_{0}^{1}dx_{1} {\int}_{0}^{1}dx_{2} {\int}_{0}^{1}dx_{3}
  {\int}_{0}^{\infty}b_{1}db_{1}  {\int}_{0}^{\infty}b_{2}db_{2}
  {\int}_{0}^{\infty}db_{3}
   \nonumber \\ & &
  {\phi}_{B_{c}}(x_{2})\, E_{f}(t_{f})\,
  H_{ef}({\alpha}_{a},{\beta}_{f},b_{1},b_{2})\,
  {\alpha}_{s}(t_{f})\, {\delta}(b_{2}-b_{3})
   \nonumber \\ & &
   \Big\{ {\phi}_{\Upsilon}^{v}(x_{1})\,
  {\phi}_{D_{s}^{\ast}}^{t}(x_{3})\, m_{2}\,m_{3}\,
   \Big[2\,m_{1}^{2}\,\bar{x}_{1}-t\,x_{2}-u\,\bar{x}_{3} \Big]
   \nonumber \\ & &
 +{\phi}_{\Upsilon}^{v}(x_{1})\,
  {\phi}_{D_{s}^{\ast}}^{v}(x_{3})\,
  4\,m_{1}^{2}\,p^{2}\,(\bar{x}_{1}-x_{2})
   \nonumber \\ & &
 -{\phi}_{\Upsilon}^{t}(x_{1})\,
   {\phi}_{D_{s}^{\ast}}^{v}(x_{3})\,
   m_{1}\,m_{b}\,s \Big\}
   \label{figfL-LL},
   \end{eqnarray}
   \begin{eqnarray}
  {\cal A}_{f,N}^{LL} &=& \frac{1}{N_{c}}
  {\int}_{0}^{1}dx_{1} {\int}_{0}^{1}dx_{2} {\int}_{0}^{1}dx_{3}
  {\int}_{0}^{\infty}b_{1}db_{1}  {\int}_{0}^{\infty}b_{2}db_{2}
  {\int}_{0}^{\infty}db_{3}
   \nonumber \\ & &
  {\phi}_{B_{c}}(x_{2})\, E_{f}(t_{f})\,
  H_{ef}({\alpha}_{a},{\beta}_{f},b_{1},b_{2})\,
  {\alpha}_{s}(t_{f})\, {\delta}(b_{2}-b_{3})
   \nonumber \\ & &
   \Big\{ {\phi}_{\Upsilon}^{V}(x_{1})\,
  {\phi}_{D_{s}^{\ast}}^{T}(x_{3})\, m_{1}\,m_{2}\,
  (s\,x_{2}+2\,m_{3}^{2}\,\bar{x}_{3}-u\,\bar{x}_{1} )
   \nonumber \\ & &
 +{\phi}_{\Upsilon}^{T}(x_{1})\,
  {\phi}_{D_{s}^{\ast}}^{V}(x_{3})\,
   m_{3}\,m_{b}\,t \Big\}
   \label{figfN-LL},
   \end{eqnarray}
   \begin{eqnarray}
  {\cal A}_{f,T}^{LL} &=& \frac{1}{N_{c}}
  {\int}_{0}^{1}dx_{1} {\int}_{0}^{1}dx_{2} {\int}_{0}^{1}dx_{3}
  {\int}_{0}^{\infty}b_{1}db_{1}  {\int}_{0}^{\infty}b_{2}db_{2}
  {\int}_{0}^{\infty}db_{3}
   \nonumber \\ & &
  {\phi}_{B_{c}}(x_{2})\, E_{f}(t_{f})\,
  H_{ef}({\alpha}_{a},{\beta}_{f},b_{1},b_{2})\,
  {\alpha}_{s}(t_{f})\, {\delta}(b_{2}-b_{3})
   \nonumber \\ & &
   \Big\{ {\phi}_{\Upsilon}^{V}(x_{1})\,
  {\phi}_{D_{s}^{\ast}}^{T}(x_{3})\, 2\,m_{1}\,m_{2}\,
  (\bar{x}_{1}-x_{2})
   \nonumber \\ & &
 -{\phi}_{\Upsilon}^{T}(x_{1})\,
  {\phi}_{D_{s}^{\ast}}^{V}(x_{3})\,
   2\,m_{3}\,m_{b} \Big\}
   \label{figfT-LL},
   \end{eqnarray}
   \begin{eqnarray}
  {\cal A}_{f,L}^{LR} &=& \frac{1}{N_{c}}
  {\int}_{0}^{1}dx_{1} {\int}_{0}^{1}dx_{2} {\int}_{0}^{1}dx_{3}
  {\int}_{0}^{\infty}b_{1}db_{1}  {\int}_{0}^{\infty}b_{2}db_{2}
  {\int}_{0}^{\infty}db_{3}
   \nonumber \\ & &
  {\phi}_{B_{c}}(x_{2})\, E_{f}(t_{f})\,
  H_{ef}({\alpha}_{a},{\beta}_{f},b_{1},b_{2})\,
  {\alpha}_{s}(t_{f})\, {\delta}(b_{2}-b_{3})
   \nonumber \\ & &
   \Big\{ {\phi}_{\Upsilon}^{v}(x_{1})\,
  {\phi}_{D_{s}^{\ast}}^{v}(x_{3})\, u\,
   \Big[ 2\,m_{2}^{2}\,x_{2}+s\,\bar{x}_{3}-t\,\bar{x}_{1} \Big]
   \nonumber \\ & &
  + {\phi}_{\Upsilon}^{v}(x_{1})\,
  {\phi}_{D_{s}^{\ast}}^{t}(x_{3})\, m_{2}\,m_{3}\,
   \Big[ t\,x_{2}+u\,\bar{x}_{3}-2\,m_{1}^{2}\,\bar{x}_{1} \Big]
   \nonumber \\ & &
 +{\phi}_{\Upsilon}^{t}(x_{1})\,m_{1}\,m_{b}\, \Big[
  {\phi}_{D_{s}^{\ast}}^{v}(x_{3})\,s
 +{\phi}_{D_{s}^{\ast}}^{t}(x_{3})\,
   4\,m_{2}\,m_{3} \Big] \Big\}
   \label{figfL-LR},
   \end{eqnarray}
   \begin{eqnarray}
  {\cal A}_{f,N}^{LR} &=& \frac{1}{N_{c}}
  {\int}_{0}^{1}dx_{1} {\int}_{0}^{1}dx_{2} {\int}_{0}^{1}dx_{3}
  {\int}_{0}^{\infty}b_{1}db_{1}  {\int}_{0}^{\infty}b_{2}db_{2}
  {\int}_{0}^{\infty}db_{3}
   \nonumber \\ & &
  {\phi}_{B_{c}}(x_{2})\, E_{f}(t_{f})\,
  H_{ef}({\alpha}_{a},{\beta}_{f},b_{1},b_{2})\,
  {\alpha}_{s}(t_{f})\, {\delta}(b_{2}-b_{3})
   \nonumber \\ & &
   \Big\{ {\phi}_{\Upsilon}^{V}(x_{1})\,
  {\phi}_{D_{s}^{\ast}}^{V}(x_{3})\, 2\, m_{1}\, m_{3}\,
   \Big[ t\,\bar{x}_{1}-2\,m_{2}^{2}\,x_{2}-s\,\bar{x}_{3} \Big]
   \nonumber \\ & &
  + {\phi}_{\Upsilon}^{V}(x_{1})\,
  {\phi}_{D_{s}^{\ast}}^{T}(x_{3})\, m_{1}\,m_{2}\,
   \Big[ u\,\bar{x}_{1}-s\,x_{2}-2\,m_{3}^{2}\,\bar{x}_{3} \Big]
   \nonumber \\ & &
 -{\phi}_{\Upsilon}^{T}(x_{1})\,m_{b}\, \Big[
  {\phi}_{D_{s}^{\ast}}^{V}(x_{3})\,m_{3}\,t
 +{\phi}_{D_{s}^{\ast}}^{T}(x_{3})\,2\,m_{2}\,u
   \Big] \Big\}
   \label{figfN-LR},
   \end{eqnarray}
   \begin{eqnarray}
  {\cal A}_{f,T}^{LR} &=& \frac{1}{N_{c}}
  {\int}_{0}^{1}dx_{1} {\int}_{0}^{1}dx_{2} {\int}_{0}^{1}dx_{3}
  {\int}_{0}^{\infty}b_{1}db_{1}  {\int}_{0}^{\infty}b_{2}db_{2}
  {\int}_{0}^{\infty}db_{3}
   \nonumber \\ & &
  {\phi}_{B_{c}}(x_{2})\, E_{f}(t_{f})\,
  H_{ef}({\alpha}_{a},{\beta}_{f},b_{1},b_{2})\,
  {\alpha}_{s}(t_{f})\, {\delta}(b_{2}-b_{3})
   \nonumber \\ & &
   \Big\{ {\phi}_{\Upsilon}^{V}(x_{1})\,
  {\phi}_{D_{s}^{\ast}}^{T}(x_{3})\, 2\, m_{1}\, m_{2}\,
   (x_{2}-\bar{x}_{1})
   \nonumber \\ & &
 +{\phi}_{\Upsilon}^{T}(x_{1})\,2\,m_{b}\, \Big[
  {\phi}_{D_{s}^{\ast}}^{T}(x_{3})\,2\,m_{2}
 -{\phi}_{D_{s}^{\ast}}^{V}(x_{3})\,m_{3} \Big] \Big\}
   \label{figfT-LR},
   \end{eqnarray}
   \begin{eqnarray}
  {\cal A}_{f,L}^{SP} &=& \frac{1}{N_{c}}
  {\int}_{0}^{1}dx_{1} {\int}_{0}^{1}dx_{2} {\int}_{0}^{1}dx_{3}
  {\int}_{0}^{\infty}b_{1}db_{1}  {\int}_{0}^{\infty}b_{2}db_{2}
  {\int}_{0}^{\infty}db_{3}
   \nonumber \\ & &
  {\phi}_{B_{c}}(x_{2})\, E_{f}(t_{f})\,
  H_{ef}({\alpha}_{a},{\beta}_{f},b_{1},b_{2})\,
  {\alpha}_{s}(t_{f})\, {\delta}(b_{2}-b_{3})
   \nonumber \\ & &
   \Big\{ {\phi}_{\Upsilon}^{t}(x_{1})\,
  {\phi}_{D_{s}^{\ast}}^{t}(x_{3})\, m_{1}\, m_{3}\,
   \Big[ t\,\bar{x}_{1}-2\,m_{2}^{2}\,x_{2}-s\,\bar{x}_{3} \Big]
   \nonumber \\ & &
  + {\phi}_{\Upsilon}^{t}(x_{1})\,
  {\phi}_{D_{s}^{\ast}}^{v}(x_{3})\, m_{1}\, m_{2}\,
   \Big[ u\,\bar{x}_{1}-s\,x_{2}-2\,m_{3}^{2}\,\bar{x}_{3} \Big]
   \nonumber \\ & &
 -{\phi}_{\Upsilon}^{v}(x_{1})\,m_{b}\, \Big[
  {\phi}_{D_{s}^{\ast}}^{v}(x_{3})\,m_{2}\,u
 +{\phi}_{D_{s}^{\ast}}^{t}(x_{3})\,m_{3}\,t \Big] \Big\}
   \label{figfL-SP},
   \end{eqnarray}
   \begin{eqnarray}
  {\cal A}_{f,N}^{SP} &=& \frac{1}{N_{c}}
  {\int}_{0}^{1}dx_{1} {\int}_{0}^{1}dx_{2} {\int}_{0}^{1}dx_{3}
  {\int}_{0}^{\infty}b_{1}db_{1}  {\int}_{0}^{\infty}b_{2}db_{2}
  {\int}_{0}^{\infty}db_{3}
   \nonumber \\ & &
  {\phi}_{B_{c}}(x_{2})\, E_{f}(t_{f})\,
  H_{ef}({\alpha}_{a},{\beta}_{f},b_{1},b_{2})\,
  {\alpha}_{s}(t_{f})\, {\delta}(b_{2}-b_{3})
   \nonumber \\ & &
   \Big\{ {\phi}_{\Upsilon}^{T}(x_{1})\, \Big[
  {\phi}_{D_{s}^{\ast}}^{V}(x_{3})\, m_{2}\, m_{3}\,
   ( t\,x_{2}+u\,\bar{x}_{3}-2\,m_{1}^{2}\,\bar{x}_{1} )
   \nonumber \\ & &
 +{\phi}_{D_{s}^{\ast}}^{T}(x_{3})\,
   \{ (s\,t-m_{2}^{2}\,u)\,x_{2}+m_{3}^{2}\,t\,\bar{x}_{3}
   -m_{1}^{2}\,s\,\bar{x}_{1} \} \Big]
   \nonumber \\ & &
 +{\phi}_{\Upsilon}^{V}(x_{1})\,m_{1}\,m_{b}\, \Big[
  {\phi}_{D_{s}^{\ast}}^{V}(x_{3})\,2\,m_{2}\,m_{3}
 +{\phi}_{D_{s}^{\ast}}^{T}(x_{3})\,s \Big] \Big\}
   \label{figfN-SP},
   \end{eqnarray}
   \begin{eqnarray}
  {\cal A}_{f,T}^{SP} &=& \frac{1}{N_{c}}
  {\int}_{0}^{1}dx_{1} {\int}_{0}^{1}dx_{2} {\int}_{0}^{1}dx_{3}
  {\int}_{0}^{\infty}b_{1}db_{1}  {\int}_{0}^{\infty}b_{2}db_{2}
  {\int}_{0}^{\infty}db_{3}
   \nonumber \\ & &
  {\phi}_{B_{c}}(x_{2})\, E_{f}(t_{f})\,
  H_{ef}({\alpha}_{a},{\beta}_{f},b_{1},b_{2})\,
  {\alpha}_{s}(t_{f})\, {\delta}(b_{2}-b_{3})
   \nonumber \\ & &
   \Big\{ {\phi}_{\Upsilon}^{T}(x_{1})\, \Big[
  {\phi}_{D_{s}^{\ast}}^{V}(x_{3})\, 2\,m_{2}\, m_{3}\,
   ( x_{2}-\bar{x}_{3} )
   \nonumber \\ & &
 +{\phi}_{D_{s}^{\ast}}^{T}(x_{3})\,
   \{ (s+t)\,x_{2}+(u-s)\,\bar{x}_{3}
   -2\,m_{1}^{2}\,\bar{x}_{1} \} \Big]
   \nonumber \\ & &
 +{\phi}_{\Upsilon}^{V}(x_{1})\,
  {\phi}_{D_{s}^{\ast}}^{T}(x_{3})\,2\,m_{1}\,m_{b} \Big\}
   \label{figfT-SP},
   \end{eqnarray}
   \begin{eqnarray}
  {\cal A}_{g,L}^{LL} &=& {\cal A}_{g,L}^{LR}\ =\
  {\int}_{0}^{1}dx_{2} {\int}_{0}^{1}dx_{3}
  {\int}_{0}^{\infty}b_{2} db_{2}
  {\int}_{0}^{\infty}b_{3} db_{3}
   \nonumber \\ & &
  {\phi}_{B_{c}}(x_{2})\, {\phi}_{D_{s}^{\ast}}^{v}(x_{3})\,
  E_{g}(t_{g})\, H_{gh}({\alpha}_{a},{\beta}_{g},b_{2},b_{3})
  \nonumber \\ & &
  {\alpha}_{s}(t_{g})\, \Big\{ m_{3}^{2}\,t+
   ( 4\,m_{1}^{2}\,p^{2}+m_{2}^{2}\,u )\,x_{2} \Big\}
   \label{figgL-LL},
   \end{eqnarray}
   \begin{eqnarray}
  {\cal A}_{g,N}^{LL} &=& {\cal A}_{g,N}^{LR}\ =\
 -{\int}_{0}^{1}dx_{2} {\int}_{0}^{1}dx_{3}
  {\int}_{0}^{\infty}b_{2} db_{2}
  {\int}_{0}^{\infty}b_{3} db_{3}
   \nonumber \\ & &
  {\phi}_{B_{c}}(x_{2})\, {\phi}_{D_{s}^{\ast}}^{V}(x_{3})\,
  E_{g}(t_{g})\, H_{gh}({\alpha}_{a},{\beta}_{g},b_{2},b_{3})
  \nonumber \\ & &
  {\alpha}_{s}(t_{g})\, m_{1}\,m_{3}\,
  \Big\{ s+2\,m_{2}^{2}\,x_{2} \Big\}
   \label{figgN-LL},
   \end{eqnarray}
   \begin{eqnarray}
  {\cal A}_{g,T}^{LL} &=& {\cal A}_{g,T}^{LR}\ =\
  {\int}_{0}^{1}dx_{2} {\int}_{0}^{1}dx_{3}
  {\int}_{0}^{\infty}b_{2} db_{2}
  {\int}_{0}^{\infty}b_{3} db_{3}
   \nonumber \\ & &
  {\phi}_{B_{c}}(x_{2})\, {\phi}_{D_{s}^{\ast}}^{V}(x_{3})\,
  E_{g}(t_{g})\, H_{gh}({\alpha}_{a},{\beta}_{g},b_{2},b_{3})
  \nonumber \\ & &
  {\alpha}_{s}(t_{g})\, 2\,m_{1}\,m_{3}
   \label{figgT-LL},
   \end{eqnarray}
   \begin{eqnarray}
  {\cal A}_{h,L}^{LL} &=& {\cal A}_{h,L}^{LR}\ =\
  {\int}_{0}^{1}dx_{2} {\int}_{0}^{1}dx_{3}
  {\int}_{0}^{\infty}b_{2} db_{2}
  {\int}_{0}^{\infty}b_{3} db_{3}
   \nonumber \\ & &
  {\phi}_{B_{c}}(x_{2})\, E_{h}(t_{h})\,
   H_{gh}({\alpha}_{a},{\beta}_{h},b_{3},b_{2})\,
  {\alpha}_{s}(t_{h})
   \nonumber \\ & &
   \Big\{ {\phi}_{D_{s}^{\ast}}^{t}(x_{3})\, \Big[
   2\,m_{2}\,m_{3}(t+u\,\bar{x}_{3})-m_{3}\,m_{b}\,t
   \Big]
   \nonumber \\ & &
  +{\phi}_{D_{s}^{\ast}}^{v}(x_{3})\,
   \Big[ (4\,m_{1}^{2}\,p^{2}+m_{3}^{2}\,t)\,\bar{x}_{3}
   + m_{2}^{2}\,u
   \nonumber \\ & & \qquad \qquad
   -2\,m_{2}\,m_{b}\,u \Big] \Big\}
   \label{fighL-LL},
   \end{eqnarray}
   \begin{eqnarray}
  {\cal A}_{h,N}^{LL} &=& {\cal A}_{h,N}^{LR}\ =\
  {\int}_{0}^{1}dx_{2} {\int}_{0}^{1}dx_{3}
  {\int}_{0}^{\infty}b_{2} db_{2}
  {\int}_{0}^{\infty}b_{3} db_{3}
   \nonumber \\ & &
  {\phi}_{B_{c}}(x_{2})\, E_{h}(t_{h})\,
   H_{gh}({\alpha}_{a},{\beta}_{h},b_{3},b_{2})\,
  {\alpha}_{s}(t_{h})
   \nonumber \\ & &
   \Big\{ {\phi}_{D_{s}^{\ast}}^{V}(x_{3})\, m_{1}\,m_{3}\,\Big[
   4\,m_{2}\,m_{b}-2\,m_{2}^{2}-s\,\bar{x}_{3} \Big]
   \nonumber \\ & &
  +{\phi}_{D_{s}^{\ast}}^{T}(x_{3})\,m_{1}\,
   \Big[ m_{b}\,s-2\,m_{2}\,s-4\,m_{2}\,m_{3}^{2}\,\bar{x}_{3}
   \Big] \Big\}
   \label{fighN-LL},
   \end{eqnarray}
   \begin{eqnarray}
  {\cal A}_{h,T}^{LL} &=& {\cal A}_{h,T}^{LR}\ =\
  {\int}_{0}^{1}dx_{2} {\int}_{0}^{1}dx_{3}
  {\int}_{0}^{\infty}b_{2} db_{2}
  {\int}_{0}^{\infty}b_{3} db_{3}
   \nonumber \\ & & 2\,m_{1}\,
  {\phi}_{B_{c}}(x_{2})\, E_{h}(t_{h})\,
   H_{gh}({\alpha}_{a},{\beta}_{h},b_{3},b_{2})\,
  {\alpha}_{s}(t_{h})
   \nonumber \\ & &
  \Big\{ {\phi}_{D_{s}^{\ast}}^{T}(x_{3})\,
  (2\,m_{2}-m_{b}) - {\phi}_{D_{s}^{\ast}}^{V}(x_{3})\,m_{3}\,\bar{x}_{3} \Big\}
   \label{fighT-LL},
   \end{eqnarray}
  where $\bar{x}_{i}$ $=$ $1$ $-$ $x_{i}$;
  variable $x_{i}$ is the longitudinal momentum fraction
  of the valence quark;
  $b_{i}$ is the conjugate variable of the
  transverse momentum $k_{iT}$;
  and ${\alpha}_{s}(t)$ is the QCD coupling at the
  scale of $t$.

  The function $H_{i}$ are defined as follows \cite{plb752}.
   \begin{eqnarray}
   H_{ab}({\alpha}_{e},{\beta},b_{i},b_{j})
   &=& K_{0}(\sqrt{-{\alpha}_{e}}b_{i})
   \Big\{ {\theta}(b_{i}-b_{j})
   K_{0}(\sqrt{-{\beta}}b_{i})
   I_{0}(\sqrt{-{\beta}}b_{j})
   + (b_{i}{\leftrightarrow}b_{j}) \Big\}
   \label{hab}, \\
   H_{cd}({\alpha}_{e},{\beta},b_{2},b_{3}) &=&
   \Big\{ {\theta}(-{\beta}) K_{0}(\sqrt{-{\beta}}b_{3})
  +\frac{{\pi}}{2} {\theta}({\beta}) \Big[
   iJ_{0}(\sqrt{{\beta}}b_{3})
   -Y_{0}(\sqrt{{\beta}}b_{3}) \Big] \Big\}
   \nonumber \\ &{\times}&
   \Big\{ {\theta}(b_{2}-b_{3})
   K_{0}(\sqrt{-{\alpha}_{e}}b_{2})
   I_{0}(\sqrt{-{\alpha}_{e}}b_{3})
   + (b_{2}{\leftrightarrow}b_{3}) \Big\}
   \label{hcd}, \\
   H_{ef}({\alpha}_{a},{\beta},b_{1},b_{2}) &=&
   \Big\{ {\theta}(-{\beta}) K_{0}(\sqrt{-{\beta}}b_{1})
  +\frac{{\pi}}{2} {\theta}({\beta}) \Big[
   iJ_{0}(\sqrt{{\beta}}b_{1})
   -Y_{0}(\sqrt{{\beta}}b_{1}) \Big] \Big\}
   \nonumber \\ & & \!\!\!\!\!\!\!\!\!\!\!\!\!\!\!\!\!\!\!\!\!\!\!\!
  {\times} \frac{{\pi}}{2} \Big\{ {\theta}(b_{1}-b_{2})
   \Big[ iJ_{0}(\sqrt{{\alpha}_{a}}b_{1})
   -Y_{0}(\sqrt{{\alpha}_{a}}b_{1}) \Big]
   J_{0}(\sqrt{{\alpha}_{a}}b_{2})
   + (b_{1}{\leftrightarrow}b_{2}) \Big\}
   \label{hef}, \\
  H_{hg}({\alpha}_{a},{\beta},b_{i},b_{j}) &=&
  \frac{{\pi}^{2}}{4}
  \Big\{ iJ_{0}(\sqrt{{\alpha}_{a}}b_{j})
   -Y_{0}(\sqrt{{\alpha}_{a}}b_{j}) \Big\}
   \nonumber \\ &{\times}&
   \Big\{ {\theta}(b_{i}-b_{j})
   \Big[ iJ_{0}(\sqrt{{\beta}}b_{i})
   -Y_{0}(\sqrt{{\beta}}b_{i}) \Big]
   J_{0}(\sqrt{{\beta}}b_{j})
   + (b_{i}{\leftrightarrow}b_{j}) \Big\}
   \label{hgh},
   \end{eqnarray}
  where $J_{0}$ and $Y_{0}$ ($I_{0}$ and $K_{0}$) are the
  (modified) Bessel function of the first and second kind,
  respectively;
  ${\alpha}_{e}$ (${\alpha}_{a}$) is the
  gluon virtuality of the emission (annihilation)
  topological diagrams;
  the subscript of the quark virtuality ${\beta}_{i}$
  corresponds to the indices of Fig.\ref{fig:fey}.
  The definition of the particle virtuality is
  listed as follows \cite{plb752}.
   \begin{eqnarray}
  {\alpha}_{e} &=& \bar{x}_{1}^{2}\,m_{1}^{2}
                +  \bar{x}_{2}^{2}\,m_{2}^{2}
                -  \bar{x}_{1}\,\bar{x}_{2}\,t
   \label{gluon-q2-e}, \\
  {\alpha}_{a} &=& x_{2}^{2}\,m_{2}^{2}
                +  \bar{x}_{3}^{2}\,m_{3}^{2}
                +  x_{2}\,\bar{x}_{3}\,s
   \label{gluon-q2-a}, \\
  {\beta}_{a} &=& m_{1}^{2} - m_{b}^{2}
               +  \bar{x}_{2}^{2}\,m_{2}^{2}
               -  \bar{x}_{2}\,t
   \label{beta-fa}, \\
  {\beta}_{b} &=& m_{2}^{2} - m_{c}^{2}
               +  \bar{x}_{1}^{2}\,m_{1}^{2}
               -  \bar{x}_{1}\,t
   \label{beta-fb}, \\
  {\beta}_{c} &=& x_{1}^{2}\,m_{1}^{2}
               +  x_{2}^{2}\,m_{2}^{2}
               +  \bar{x}_{3}^{2}\,m_{3}^{2}
   \nonumber \\ &-&
                  x_{1}\,x_{2}\,t
               -  x_{1}\,\bar{x}_{3}\,u
               +  x_{2}\,\bar{x}_{3}\,s
   \label{beta-fc}, \\
  {\beta}_{d} &=& x_{1}^{2}\,m_{1}^{2}
               +  x_{2}^{2}\,m_{2}^{2}
               +  x_{3}^{2}\,m_{3}^{2}
               -  m_{c}^{2}
    \nonumber \\ &-&
                  x_{1}\,x_{2}\,t
               -  x_{1}\,x_{3}\,u
               +  x_{2}\,x_{3}\,s
   \label{beta-fd}, \\
  {\beta}_{e} &=& x_{1}^{2}\,m_{1}^{2}
               +  x_{2}^{2}\,m_{2}^{2}
               + \bar{x}_{3}^{2}\,m_{3}^{2}
               -  m_{b}^{2}
   \nonumber \\ &-&
                  x_{1}\,x_{2}\,t
               -  x_{1}\,\bar{x}_{3}\,u
               +  x_{2}\,\bar{x}_{3}\,s
   \label{beta-fe}, \\
  {\beta}_{f} &=& \bar{x}_{1}^{2}\,m_{1}^{2}
               +  x_{2}^{2}\,m_{2}^{2}
               +  \bar{x}_{3}^{2}\,m_{3}^{2}
               -  m_{b}^{2}
   \nonumber \\ &-&
                  \bar{x}_{1}\,x_{2}\,t
               -  \bar{x}_{1}\,\bar{x}_{3}\,u
               +  x_{2}\,\bar{x}_{3}\,s
   \label{beta-ff}, \\
  {\beta}_{g} &=& x_{2}^{2}\,m_{2}^{2}
               +  m_{3}^{2}
               +  x_{2}\,s
   \label{beta-fg}, \\
  {\beta}_{h} &=& \bar{x}_{3}^{2}\,m_{3}^{2}
               +  m_{2}^{2}
               +  \bar{x}_{3}\,s
               - m_{b}^{2}
   \label{beta-fh}.
   \end{eqnarray}

  The typical scale $t_{i}$ and the Sudakov factor $E_{i}$
  are defined as follows, where the subscript $i$ corresponds
  to the indices of Fig.\ref{fig:fey}.
   \begin{eqnarray}
   t_{a(b)} &=& {\max}(\sqrt{-{\alpha}_{e}},\sqrt{-{\beta}_{a(b)}},1/b_{1},1/b_{2})
   \label{tab}, \\
   t_{c(d)} &=& {\max}(\sqrt{-{\alpha}_{e}},\sqrt{{\vert}{\beta}_{c(d)}{\vert}},1/b_{2},1/b_{3})
   \label{tcd}, \\
   t_{e(f)} &=& {\max}(\sqrt{{\alpha}_{a}},\sqrt{{\vert}{\beta}_{e(f)}{\vert}},1/b_{1},1/b_{2})
   \label{tef}, \\
   t_{g(h)} &=& {\max}(\sqrt{{\alpha}_{a}},\sqrt{{\beta}_{g(h)}},1/b_{2},1/b_{3})
   \label{tgh},
   \end{eqnarray}
   \begin{equation}
   E_{i}(t) =
   \left\{ \begin{array}{lll}
  {\exp}\{ -S_{{\Upsilon}(1S)}(t)-S_{B_{c}}(t) \}, &~& i=a,b \\
  {\exp}\{ -S_{{\Upsilon}(1S)}(t)-S_{B_{c}}(t)-S_{D_{s}^{\ast}}(t) \}, & & i=c,d,e,f \\
  {\exp}\{ -S_{B_{c}}(t)-S_{D_{s}^{\ast}}(t) \}, & & i=g,h
   \end{array} \right.
   \label{sudakov-exp},
   \end{equation}
   \begin{eqnarray}
   S_{{\Upsilon}(1S)}(t) &=&
   s(x_{1},p_{1}^{+},1/b_{1})
  +2{\int}_{1/b_{1}}^{t}\frac{d{\mu}}{\mu}{\gamma}_{q}
   \label{sudakov-bb}, \\
   S_{B_{c}}(t) &=&
   s(x_{2},p_{2}^{+},1/b_{2})
  +2{\int}_{1/b_{2}}^{t}\frac{d{\mu}}{\mu}{\gamma}_{q}
   \label{sudakov-bc}, \\
   S_{D_{s}^{\ast}}(t) &=&
   s(x_{3},p_{3}^{+},1/b_{3})
  +2{\int}_{1/b_{3}}^{t}\frac{d{\mu}}{\mu}{\gamma}_{q}
   \label{sudakov-ds},
   \end{eqnarray}
  where ${\gamma}_{q}$ $=$ $-{\alpha}_{s}/{\pi}$ is the
  quark anomalous dimension;
  the explicit expression of $s(x,Q,1/b)$ can be found in
  the appendix of Ref.\cite{pqcd1}.
  \end{appendix}

  
  \end{document}